\documentclass[longauth]{aa}  

\usepackage{graphicx}
\usepackage{txfonts}
\usepackage{multirow}
\usepackage{gensymb}
\usepackage{color}
\usepackage{hyperref}
\hypersetup{breaklinks=true,colorlinks=true,urlcolor=blue, linkcolor=blue, citecolor=blue}

\begin{document}

   \title{The SPHERE view of the Taurus star-forming region \thanks{Based on observations collected at the European Southern Observatory under the following ESO programs: 096.C-0248, 098.C-0760, 0100.C-0408, 0100.C-0452, 0101.C-0867, 0102.C-0453, 0102.C-0916, 0104.C-0122, 106.21HJ.001, 108.22EE.001, 198.C-0209, 1104.C-0415}}

   \subtitle{The full census of planet-forming disks with GTO and DESTINYS programs}

   \author{A. Garufi\inst{\ref{inst:arcetri}}
          \and
          C. Ginski\inst{\ref{inst:galway}}
          \and
          R.\,G. van~Holstein\inst{\ref{inst:esosantiago}} 
          \and M. Benisty\inst{\ref{inst:nice}, \ref{inst:grenoble}}
          \and C.\,F. Manara\inst{\ref{inst:ESO}}
          \and S. P{\'e}rez\inst{\ref{inst:yems}, \ref{inst:usach}}
          \and P. Pinilla\inst{\ref{inst:ucl}}
          \and Á. Ribas\inst{\ref{inst:cambridge}}
          \and \\ P. Weber\inst{\ref{inst:yems}, \ref{inst:usach}}
          \and J. Williams\inst{\ref{inst:hawaii}}
          \and L. Cieza\inst{\ref{inst:portales}, \ref{inst:yems}}
          \and C. Dominik\inst{\ref{inst:amsterdam}}
          \and S. Facchini\inst{\ref{inst:milano}}
          \and J. Huang\inst{\ref{inst:columbia}}
          \and A. Zurlo\inst{\ref{inst:portales}, \ref{inst:yems}}
          \and J. Bae\inst{\ref{inst:gainesville}}
          \and \\ J. Hagelberg\inst{\ref{inst:geneva}}
          \and Th. Henning\inst{\ref{inst:heidelberg}}
          \and M.\,R. Hogerheijde\inst{\ref{inst:leiden}, \ref{inst:amsterdam}}
          \and M. Janson\inst{\ref{inst:stockholm}}
          \and F. M{\'e}nard\inst{\ref{inst:grenoble}}
          \and S. Messina \inst{\ref{inst:catania}}
          \and M.\,R. Meyer\inst{\ref{inst:michigan}}
          \and \\ C. Pinte
          \inst{\ref{inst:monash}, \ref{inst:grenoble}}
          \and S.\,P. Quanz\inst{\ref{inst:eth}}
          \and E. Rigliaco\inst{\ref{inst:padova}}
          \and V. Roccatagliata\inst{\ref{inst:arcetri}}
          \and H.\,M. Schmid\inst{\ref{inst:eth}}
          \and J. Szul{\'a}gyi\inst{\ref{inst:eth}}
          \and \\ R. van Boekel\inst{\ref{inst:heidelberg}}
          \and Z. Wahhaj\inst{\ref{inst:esosantiago}}
          \and J. Antichi\inst{\ref{inst:arcetri}}
          \and A. Baruffolo\inst{\ref{inst:padova}}
          \and T. Moulin\inst{\ref{inst:grenoble}}
          }
          
   \institute{ INAF, Osservatorio Astrofisico di Arcetri, Largo Enrico Fermi 5, I-50125 Firenze, Italy\\
              \email{antonio.garufi@inaf.it} \label{inst:arcetri}
         \and Centre for Astronomy, Dept. of Physics, National University of Ireland Galway, University Road, Galway H91 TK33, Ireland \label{inst:galway}    
         \and European Southern Observatory, Alonso de C\'{o}rdova 3107, Casilla 19001, Vitacura, Santiago, Chile \label{inst:esosantiago}
         \and Laboratoire Lagrange, Universit\'{e} C\^{o}te d'Azur, Observatoire de la C\^{o}te d'Azur, CNRS, Nice Cedex 4, France  \label{inst:nice}
         \and Univ. Grenoble Alpes, CNRS, IPAG, 38000 Grenoble, France \label{inst:grenoble}
         \and European Southern Observatory (ESO), Garching bei M{\"u}nchen, Germany \label{inst:ESO}
         \and Millennium Nucleus on Young Exoplanets and their Moons (YEMS), Chile  \label{inst:yems}
         \and Departamento de F{\'i}sica, Universidad de Santiago de Chile, Avenida Victor Jara 3659, Santiago, Chile \label{inst:usach}
         \and Mullard Space Science Laboratory, University College London, Holmbury St Mary, Dorking, Surrey RH5 6NT, UK \label{inst:ucl}
         \and Institute of Astronomy, University of Cambridge, Madingley Road, Cambridge, CB3 0HA, UK \label{inst:cambridge}
         \and Institute for Astronomy, University of Hawaii, Hawaii, USA \label{inst:hawaii}
         \and Instituto de Estudios Astrof{\'i}sicos, Facultad de Ingenier{\'i}a y Ciencias, Universidad Diego Portales, Avenida Ej{\'e}rcito Libertador 441, Santiago, Chile \label{inst:portales}
         \and Anton Pannekoek Institute for Astronomy, University of Amsterdam, Amsterdam, The Netherlands \label{inst:amsterdam}
         \and Dipartimento di Fisica, Universit\`{a} degli Studi di Milano, Via Celoria 16, 20133 Milano, Italy \label{inst:milano}
         \and Department of Astronomy, Columbia University, 538 W. 120th Street, Pupin Hall, New York, NY, United States of America \label{inst:columbia}
         \and Department of Astronomy, University of Florida, Gainesville, FL 32611, USA \label{inst:gainesville} 
         \and Observatoire de Gen{\`e}ve, Universit{\'e} de Gen{\`e}ve, 51 Ch. des Maillettes, 1290 Sauverny, Switzerland \label{inst:geneva}
         \and Max Planck Institute for Astronomy, Konigstuhl 17, D-69117 Heidelberg, Germany \label{inst:heidelberg}
         \and Leiden Observatory, Leiden University, PO Box 9513, 2300 RA Leiden, The Netherlands \label{inst:leiden}
         \and Department of Astronomy, Stockholm University, 106 91 Stockholm, Sweden \label{inst:stockholm}
         \and INAF - Osservatorio Astrofisico di Catania, Via S.Sofia 78, I-95123 Catania, Italy \label{inst:catania}
         \and Department of Astronomy, The University of Michigan West Hall 323, 1085 S. University Avenue  Ann Arbor, MI 48109 \label{inst:michigan}
         \and School of Physics and Astronomy, Monash University, Vic 3800, Australia \label{inst:monash}
         \and ETH Z\"urich, Department of Physics, Wolfgang-Pauli-Strasse 27, CH-8093, Z\"urich, Switzerland \label{inst:eth}
         \and Osservatorio Astronomico di Padova, Vicolo dell’Osservatorio 5, 35122 Padova, Italy \label{inst:padova}
         }

   \date{Received -; accepted -}
   
  \abstract
   {The sample of planet-forming disks observed by high-contrast imaging campaigns over the last decade is mature enough to enable the demographical analysis of individual star-forming regions. We present the full census of Taurus sources with VLT/SPHERE polarimetric images available. The whole sample sums up to 43 targets (of which 31 have not been previously published) corresponding to one-fifth of the Class II population in Taurus and about half of such objects that are observable. A large fraction of the sample is apparently made up of isolated faint disks (equally divided between small and large self-shadowed disks). Ambient signal is visible in about one-third of the sample. This probes the interaction with the environment and with companions or the outflow activity of the system. The central portion of the Taurus region almost exclusively hosts faint disks, while the periphery also hosts bright disks interacting with their surroundings. The few bright disks are found around apparently older stars. The overall picture is that the Taurus region is in an early evolutionary stage of planet formation. Yet, some objects are discussed individually, as in an intermediate or exceptional stage of the disk evolution. This census provides a first benchmark for the comparison of the disk populations in different star forming regions.}

   \keywords{protoplanetary disks -- techniques: polarimetric -- star-forming regions: Taurus
               }

 \titlerunning{The SPHERE view of the Taurus star-forming region}
    \authorrunning{Garufi et al.}

   \maketitle
%

\section{Introduction}
The study of the planet formation is hampered by the paucity of detectable protoplanets in their natal disks. The other most direct vehicle to observationally determine how planets form are resolved maps of the planet-forming disk itself, and in particular of the (evolving) morphological features that are connected to the interaction with embedded planets. This approach has been successfully carried out over the last few years thanks to the near-infrared (NIR) high-contrast imaging from the ground \citep[see review by][]{Benisty2023} and the high-resolution imaging granted by the Atacama Large (sub)Millimeter Array \citep[ALMA; see the review by][]{Andrews2020}. While the ALMA community carried out demographical studies of entire star-forming regions at moderate resolution \citep[e.g.,][]{Pascucci2016, Ansdell2016}, such an effort by the NIR community struggled to take off, mostly because of the longer minimum exposure times required on the telescopes.

About one decade after the onset of the latest-generation instruments offering high-contrast imaging (such as the Gemini Planet Imager, GPI, \citealt{Macintosh2014}, and the Spectro-Polarimetric High-contrast Exoplanet REsearch SPHERE, \citealt{Beuzit2019}), the community is finally in the condition to lead demographical studies on large samples of young stellar objects. This effort has been made possible in particular by systematic programs such as the SEEDS at Subaru \citep[e.g.,][]{Hashimoto2012, Mayama2012}, the Guaranteed Time Observation (GTO) program at SPHERE \citep[e.g.,][]{Garufi2016, Ginski2016}, DARTTS at SPHERE \citep{Avenhaus2018, Garufi2020b}, LIGHTS at GPI \citep[e.g.,][]{Laws2020, Rich2022}, and more recently DESTINYS at SPHERE \citep[e.g.,][]{Ginski2021, Ginski2022}. 

In particular,  the Disk Evolution Study Through Imaging of Nearby Young Stars (DESTINYS) is an ESO large program to obtain deep, polarized intensity images of 85 young stellar objects in all nearby star forming regions with the goal of alleviating the bias on the available sample and constraining the disk evolution over a wide range of ages. Thanks to DESTINYS and the previous programs, the demography of disks can now be studied through a significantly large and reasonably unbiased sample. A first taxonomical analysis of the full sample available at the time (58 targets) was performed by \citet{Garufi2018}, which identified six main classes of disk types based on NIR images. More recently, \citet{Rich2022} analyzed the 44 sources in LIGHTS to significantly enlarge the stellar parameter space to higher masses. Furthermore, \citet{Garufi2022b} examined a sample of 15 faint disks to demonstrate the high occurrence of self-shadowed disks, where the inner portion of the disk casts a shadow on the outer disk that is observable by the current generation of telescopes.

In this work, we took a step forward in the study of disk evolution by moving the focus of our objective from the specific type of objects or observing program to the population of young stars in a specific star-forming region. In parallel with the work by Ginski et al.\ on Chamaeleon and by Valegard et al.\ on Orion, we collected the entirety of the SPHERE dataset concerning the Taurus region that has been taken over the last decade with SPHERE/IRDIS \citep{Dohlen2008} in dual-beam polarimetric imaging \citep[DPI,][]{deBoer2020, vanHolstein2020} mode by the SPHERE GTO (PI: Beuzit), by the large program DESTINYS (PI: Ginski), and by individual open-time programs. The final sample that we built is composed of 43 targets (31 of which have not previously been published), making it the largest sample of high-contrast images from an individual region ever studied as a whole.

The Taurus-Auriga complex is one of the most studied nearby \citep[128--198 pc,][]{Galli2018}, low-mass, star-forming regions. It hosts a few hundred young stellar objects that are distributed in small kinematical groups \citep{Esplin2019, Joncour2018}. All Taurus members appear coeval \citep[1--3 Myr,][]{Luhman2023} and yet they are in very diverse evolutionary stages, ranging from embedded protostars to disk-free stars \citep{Kenyon2008}. Large-scale maps of the dust distribution revealed that Taurus lies on the near hemispheric side of the so-called Perseus-Taurus shell \citep{Bialy2020}, which suggests that the whole cloud may have been formed via compression of material between the expanding Perseus-Taurus bubble and the local bubble \citep{Soler2023}. 

For a long time, the large sky area and the low density of Taurus hindered the determination of the individual membership. The \textit{Gaia} mission \citep{Gaia2016} is now significantly reducing contamination from field stars by providing kinematical data on all stars as faint as 20 mag in the $G$ waveband, corresponding to a stellar mass of 0.05 M$_\odot$ at the distance of Taurus \citep{Luhman2023}. Several authors have employed the \textit{Gaia} data to constrain the morphological and kinematical properties of Taurus \citep[e.g.,][]{Galli2019, Roccatagliata2020, Krolikowski2021}. In this work, we made use of the membership catalog created by \citet{Luhman2023} based on the latest \textit{Gaia} release \citep[DR3,][]{Gaia2023}. We focused on the specific evolutionary stage of Class II objects \citep{Lada1987} as these stars still host a gas-rich planet-forming disk, but no longer possess any dusty envelope that obscures the optical light from the star.

The paper is organized as follows. In Sect.\,\ref{sec:sample}, we census the spatial distribution of the available SPHERE sample and calculate the stellar and disk properties. In Sect.\,\ref{sec:Observations}, we describe the observations and data reduction. The results from the SPHERE images are given in Sect.\,\ref{sec:Results}, while those from the comparison with literature ALMA images are given in Sect.\,\ref{sec:ALMA}. Finally, we discuss and summarize our results in Sects.\,\ref{Discussion} and \ref{Summary}.

\section{Sample} \label{sec:sample}
The sample studied in this work consists of the 43 sources in Taurus with available VLT/SPHERE DPI images (see Sect.\,\ref{sec:Observations}). In this section, we discuss their distribution in space and in stellar and disk properties with the intent to set a background for the SPHERE observations described in Sect.\,\ref{sec:Results}.

\subsection{Spatial distribution of the sample} \label{sec:spatial_distribution}
\begin{figure*}
  \centering
 \includegraphics[width=16.5cm]{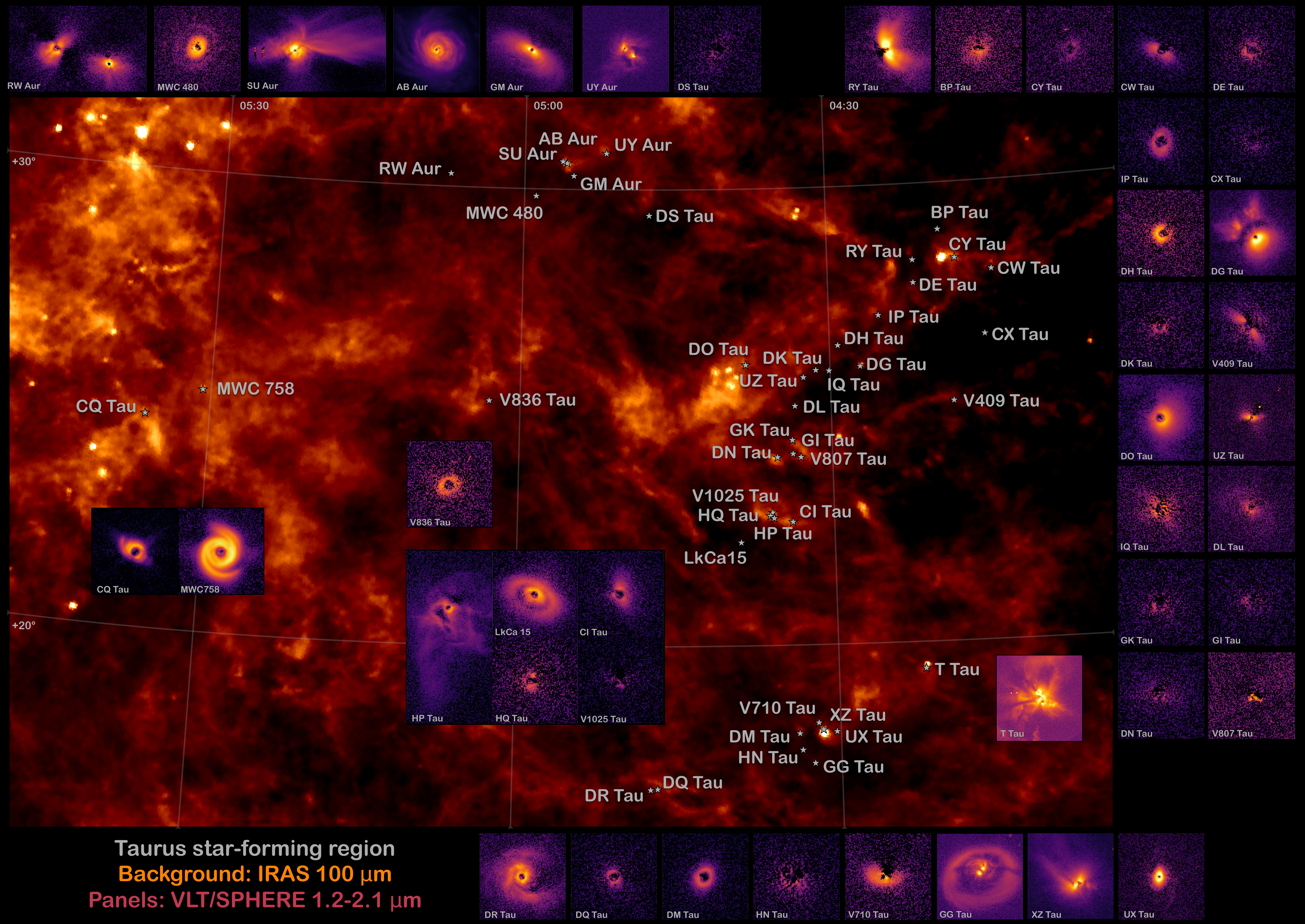} 
      \caption{Spatial distribution of sample. The background image is the Taurus region at 100 $\mu$m (Band 4) from the Improved Reprocessing of the IRAS Survey (IRIS). The SPHERE images in the $J$, $H$, or $K$ bands (see Table \ref{table:disk_sample}) are shown in the inset panels with spatial and flux scales optimized for better representation (whereas the actual scales can be seen in Fig.\ref{fig:Imagery}).}
          \label{fig:Taurus_map}
  \end{figure*}

Figure \ref{fig:Taurus_map} shows the spatial distribution of the 43 sources studied in this work. A quick look at their distribution already hints toward their membership to different groups. Following the kinematical classification by \citet{Luhman2023} based on \textit{Gaia} DR3 \citep{Gaia2023}, our targets belong to 11 different groups and one association near Taurus (namely HD35187, which hosts CQ Tau and MWC758). The groups that are most represented by our sample are L1524/L1529/B215 (10 members) and L1517 (7 members), while in five cases we only have one member. The membership and the distance from \textit{Gaia} for the entire sample are shown in Appendixes \ref{appendix:distribution} and \ref{appendix:sample}. With some notable exceptions (see below), the distance of each source is close to the averaged distance of the respective group, spanning from the 130 pc of L1495/B209 to the 196 pc of L1558. 

The goodness of the \textit{Gaia} astrometric observation used in this work is commonly evaluated through the renormalized unit weight error (RUWE). A single star with a good astrometric solution has a RUWE around 1.0, or smaller than 1.4 in any case. Interestingly, only 19 of our 43 sources have a RUWE smaller than 1.4, and these are all single stars (see Sect.\,\,\ref{sec:stellar_properties}). \citet{Fitton2022} proposed adopting a higher threshold (2.5) to evaluate the goodness of the astrometric solution on disk-bearing stars. With this threshold, 29 of our 43 sources would have a good solution. Eight of our targets have a RUWE larger than four, meaning that their astrometric solution is too poor to lend credence to the calculated distance. In one of these eight cases, GG Tau A, we used the distance of a nearby bound companion (GG Tau B) with a good solution. The other seven cases are marked in Table \ref{table:stellar_sample}. Interestingly, among these seven cases we find the aforementioned discrepancies between the individual distance and the averaged distance of the respective group (e.g., RW Aur at 183 pc from the group L1517, at 156 pc, or V807 Tau at 184 pc from L1524, lying at 128 pc), pointing to an erroneous parallax solution. 

\subsection{This sample within the whole Taurus population}
Based on \textit{Gaia} DR3, \citet{Luhman2023} identified 532 Taurus members. From the nearly identical sample by \citet{Esplin2019}, approximately 200 Taurus members have an SED indicative of a full Class II disk around the star, or $\sim$220 including transition and evolved disks (with most of the remaining sources being Class III). Given the sample size of this work, we conclude that about one-fifth of the Taurus planet-forming disks have been probed by direct NIR imaging.

The main technical limit toward the high-contrast imaging of Class II objects within 200 pc is the stellar brightness at optical wavelengths needed to drive the adaptive-optics system at the telescope. For SPHERE, this limit is set to an $R$ waveband magnitude of 13 (although an optimal performance is only expected up to 11). From the whole sample by \citet{Esplin2019}, 187 targets show a \textit{Gaia $G_{\rm RP}$} passband\footnote{The \textit{Gaia $G_{\rm RP}$} passband is wider in wavelength than the Johnson $R$. The exact conversion depends on the stellar color, but for our exercise it is not necessary as the difference is relatively small (1 mag at most).} magnitude smaller than 13. Of these, 82 are classified as either Class II or transition-disk objects. Therefore, about a half\footnote{40$\div$82=49\%, where 40 is our sample size (43) minus CQ Tau and MWC758 that are formally not in Taurus (see Sect.\,\ref{sec:spatial_distribution}), and V1025 Tau that is classified as Class III (see Sect.\,\ref{sec:trends_excess}).} of the planet-forming disks in Taurus that are potentially observable with direct NIR imaging have been surveyed. However, if we focus on the 49 Class II targets with \textit{Gaia $G_{\rm RP}$} smaller than 12 (for which the adaptive-optics performance is at worst suboptimal), then more than 75\% (that is, 37) of the possible targets have been observed. The distribution of all Taurus sources, of all Class II sources, and of all the observable Class II sources is shown in Appendix \ref{appendix:distribution}. 

The whole Taurus sample is limited in both distance and age. Therefore, the cutoff in $R$ magnitude on the available sample is reflected in a cutoff on the stellar mass only and can be set to approximately 0.4 M$_{\odot}$. This value is in fact the lowest limit of our sample (see Sect.\,\ref{sec:stellar_properties}). The only other selection criterion that may hold is an IR excess indicative of a class II object. We can therefore conclude that our sample is unbiased and moderately complete within the Taurus stars more massive than 0.4 M$_{\odot}$ hosting a full or a transition disk. However, it remains completely blind to the entire population of lower-mass stars with a disk (that is, based on \citealt{Esplin2019}, approximately two-thirds of the whole sample of stars with a disk).

\subsection{Stellar properties} \label{sec:stellar_properties}
The 43 sources of this work are divided into three A-, one \mbox{F-,} three G-, 19 K-, and 17 M-type stars. To our knowledge, thirteen of the stars are part of multiple systems, with one spectroscopic binary (DQ Tau), five systems with close companions (less than 1\arcsec), and seven with further companions. This fraction of multiple stars from our sample (30\%) is mildly lower than the overall fraction determined for solar-type stars \citep[$\sim$50\%,][]{Raghavan2010} and significantly lower than that constrained for Taurus \citep[$\sim$70\%,][]{Kraus2011}. This is not necessarily a bias since close companions can shorten the disk lifetime or alter their morphology \citep[see][and references therein]{Offner2023}.

For each target, we retrieved the effective temperature $T_{\rm eff}$ from \citet{Lopez_Valdivia2021} and \citet{Gangi2022} and constructed the spectral energy distribution (SED) through \textit{VizieR}\footnote{\url{http://vizier.cds.unistra.fr/viz-bin/VizieR}}. We then measured the stellar luminosity $L_*$ by means of a PHOENIX model of the stellar photosphere \citep{Hauschildt1999} scaled to the $Gaia$ distance and to the $R$ magnitude de-reddened by the extinction $A_{\rm V}$ calculated from the $V$, $R$, and $I$ wavebands. Error bars on the stellar luminosity are propagated from $d$, $A_{\rm V}$ (20\%), and $T_{\rm eff}$ ($\pm$100 K). Finally, we constrained the stellar mass $M_*$ and age $t$ in the current framework of isochrones through a set of pre-main-sequence tracks \citep{Siess2000, Bressan2012, Baraffe2015, Choi2016}, including the magnetic tracks by \citet{Feiden2016} that would contrast the tendency of nonmagnetic tracks to underestimate stellar mass and age \citep{Hillenbrand2004, AsensioTorres2019}. The age interval shown in Table \ref{table:stellar_sample} reflects the variety of values suggested by the different tracks. Some more details on the extraction of the SED and the calculation of the stellar properties are given in Appendix \ref{appendix:sample}.

The calculated stellar masses of the individual stellar components span from 2.5 M$_{\odot}$ (SU Aur) to 0.4 M$_{\odot}$ (8 sources) with ten intermediate-mass stars ($\geq$ 1.5 M$_{\odot}$), nine solar-mass stars ($\geq$ 0.8 M$_{\odot}$), and 36 low-mass stars. The age of the majority of the sample is younger than 3 Myr, which is in line with the known age of the region. Sources that look older from the isochrones are therefore interesting cases that are described in Sect.\,\ref{sec:Results}. Here, we only point out that the two oldest sources in the sample (CQ Tau and MWC758, 12--14 Myr) are formally not part of Taurus but rather of a nearby association with an estimated age of 18$\pm$4 Myr \citep{Luhman2023}. As for the other sources, the limited statistics and the uncertainty involved prevent us from determining any significant trend between the age of the individual groups by \citet{Luhman2023} and the age that we inferred. 

\subsection{Disk properties} \label{sec:disk_properties}
From the aforementioned SEDs, we also extracted the IR excess and derived a crude estimate of the dust mass in the disk. For the former, we measured the integrated excess over the stellar photosphere relative to the stellar luminosity in the NIR (1.2--4.5 $\mu$m), mid-IR (MIR, 4.5--22 $\mu$m), and far-IR (FIR, 22--450 $\mu$m), as also shown in Fig.\,\ref{fig:SEDs}. For the latter, we converted the observed flux at 1.3 mm under standard assumptions \citep[optically thin emission, dust temperature of 20 K, and dust opacity by][]{Beckwith1990}. The derived dust mass should be viewed as a rough estimate due to the simplicity of these assumptions \citep[see e.g.,][]{Birnstiel2018, Zhu2019, Ribas2020, Binkert2021, Binkert2023}. The values of excess and disk mass calculated for the entire sample are shown in Table \ref{table:disk_sample}.

NIR, MIR, and FIR excesses reveal the reprocessing of the stellar photons by the hot, warm, and cold dust at the disk surface or any possible material above it. Empirically, the NIR excess of a standard, continuous disk is around 10--20\% of the stellar luminosity \citep[e.g.,][]{Garufi2022b}. Lower values point to a disk depleted in the inner regions, while higher values to an extra component of hot dust (such as uplifted material in the stellar proximity) reprocessing the stellar light. On the other hand, the FIR excess typically ranges from 5\% for cold, self-shadowed disks to $\gg$50\% for disks with heavily exposed outer disk portions. The sample of this work reflects this known variety spanning the NIR excess from 0\% (GM Aur) to 100\% (XZ Tau) and the FIR excess from 0.7\% (V1025 Tau) to 57\% (DG Tau). Similarly, the dust mass in the disk covers more than two orders of magnitude, from the 1 M$_{\oplus}$ of V1025 Tau to the 300 M$_{\oplus}$ of \mbox{GG Tau}.    

\section{SPHERE observations and data reduction} \label{sec:Observations}
By construction, all targets in the sample have SPHERE-IRDIS polarimetric images available. All of them have been observed in the $H$ broad-band filter except CQ Tau, LkCa15 (in the $J$ band), MWC758, and CY Tau (in the $K$). {Observations took place between November 2016 and December 2021. The total integration time spans from 11 to 115 minutes (see Appendix \ref{appendix:setup}).} Twelve sources have datasets published by previous authors while the other 31 are unpublished. The complete reference list (publication or observing program) can be found in Tables \ref{table:disk_sample} {and \ref{table:observations}}. Among the unpublished datasets, thirteen are from the SPHERE GTO, eleven are from DESTINYS, and seven are from various open-time programs. 

The data were reduced with the public IRDAP pipeline \citep[IRDIS Data reduction for Accurate Polarimetry;][]{vanHolstein2020} which employs a validated Mueller matrix model to correct for the instrumental polarization and crosstalk of the telescope. Although some of the datasets (e.g., all DESTINYS data) have been acquired in pupil tracking mode to produce a total-intensity image in angular differential imaging \citep[ADI,][]{Marois2006}, in this work we only focused on the polarized intensity. Our most recent work used the polarization probed by the radial Stokes parameter $Q_\phi$ following \citet{deBoer2020}:
\begin{equation}
Q_\phi=-Q\cos{2\phi}-U\sin{2\phi} 
,\end{equation}
with $\phi$ being the azimuthal angle, and $Q$ and $U$ the Stokes parameters from \citet{Stokes1851} that are extracted by IRDAP using the double-difference method \citep{vanHolstein2020}. By construction, the $Q_\phi$ parameter probes the centro-symmetric component of the polarized light and is equal to the polarized intensity $PI=\sqrt{Q^2+U^2}$ in the ideal scenario of single scattering from a face-on disk around a single star. However, several mechanisms can alter this ideal case, such as the presence of companions \citep{Weber2023}, convolution effects from even moderately inclined disks \citep{Heikamp2019}, or multiple scattering \citep{Canovas2015}. In any of these cases, a fraction of the polarized photons will move from the $Q_\phi$ parameter to the radial Stokes $U_\phi$ parameter defined, similarly to the $Q_\phi$, as:
\begin{equation}
U_\phi=+Q\sin{2\phi}-U\cos{2\phi} .
\end{equation}
In case of strong signal recorded from the $U_\phi$ parameter, we therefore adopt the $PI$ image instead of the $Q_\phi$ parameter, even though this implies squaring the noise by construction of $PI$.

The stellar polarization is measured in regions where no bound and background stars nor disks are present, and it is subsequently corrected. In some datasets, a customized procedure is adopted. This includes the visual selection of the best Stokes parameters from a grid of images with different combinations of stellar polarization, the minimization of negative signal in the $Q_\phi$ image, and the minimization of positive signal in the $U_\phi$ image. 

The final FITS file produced in our reduction follows a standard format agreed with the Subaru/HiCiao and Gemini Planet Imager communities \citep[see][]{Rich2022, Benisty2023} and contains the total intensity $I$, the Stokes parameters $Q$ and $U$, and the radial Stokes parameters $Q_\phi$ and $U_\phi$. All images are calibrated in flux by converting the pixel counts to physical units through non-coronagraphic images that are taken in parallel to the science frames. The observed count from these images is equalized to the 2MASS $JHK$ photometry of the relative star, and the conversion factor is applied to the science frames \citep[see][]{Ginski2022}. To obtain a unique value of the disk brightness for each source, we applied the method by \citet{Garufi2014b} described in detail by \citet{Garufi2022b} and \citet{Benisty2023}. The polarized-to-stellar light contrast $\alpha_{\rm pol}$ is a measurement of the fraction of photons released by the star that can reach the disk surface and are effectively scattered toward the observer. It is calculated along the disk major axis to minimize the effect of the scattering phase function and averaged between inner and outer disk radii $r_{\rm in}$ and $r_{\rm out}$ by dividing the radially integrated observed flux $S_{\rm pol}(r)$ by the stellar flux $F_*$ accounting for its dilution with the separation $r$:
\begin{equation} \label{Formula_albedo}
\alpha_{\rm pol}= \frac{1}{r_{\rm out}-r_{\rm in}} \cdot \int_{r_{\rm in}}^{r_{\rm out}} S_{\rm pol}(r)\cdot \frac{4\pi r^2}{F_*} dr .
\end{equation}

{The adoption of the $\alpha_{\rm pol}$ contrast also allows us to minimize the impact of the different filters and integration times $t_{\rm int}$ of the observations in this work. On the one hand, the $\alpha_{\rm pol}$ in the J or K bands (where we have two targets in each) is not expected to change from that in the H band unless the dust grains in the disk have a tremendously blue or red color. On the other hand, the $\alpha_{\rm pol}$ is not expected to significantly scale with the integration time while the uncertainty on $\alpha_{\rm pol}$ might have a mild dependence on $t_{\rm int}$ (as is described in Appendix \ref{appendix:setup}).}

\section{Results from the SPHERE images} \label{sec:Results}
In this section, we give an overview of the SPHERE images and report on their spatial distribution in Taurus and the correlation with the stellar and disk  properties introduced in Sect.\,\ref{sec:sample}. A brief description of the individual objects is given in Appendix \ref{appendix:individual}.   

\begin{figure*}
  \centering
 \includegraphics[width=17cm]{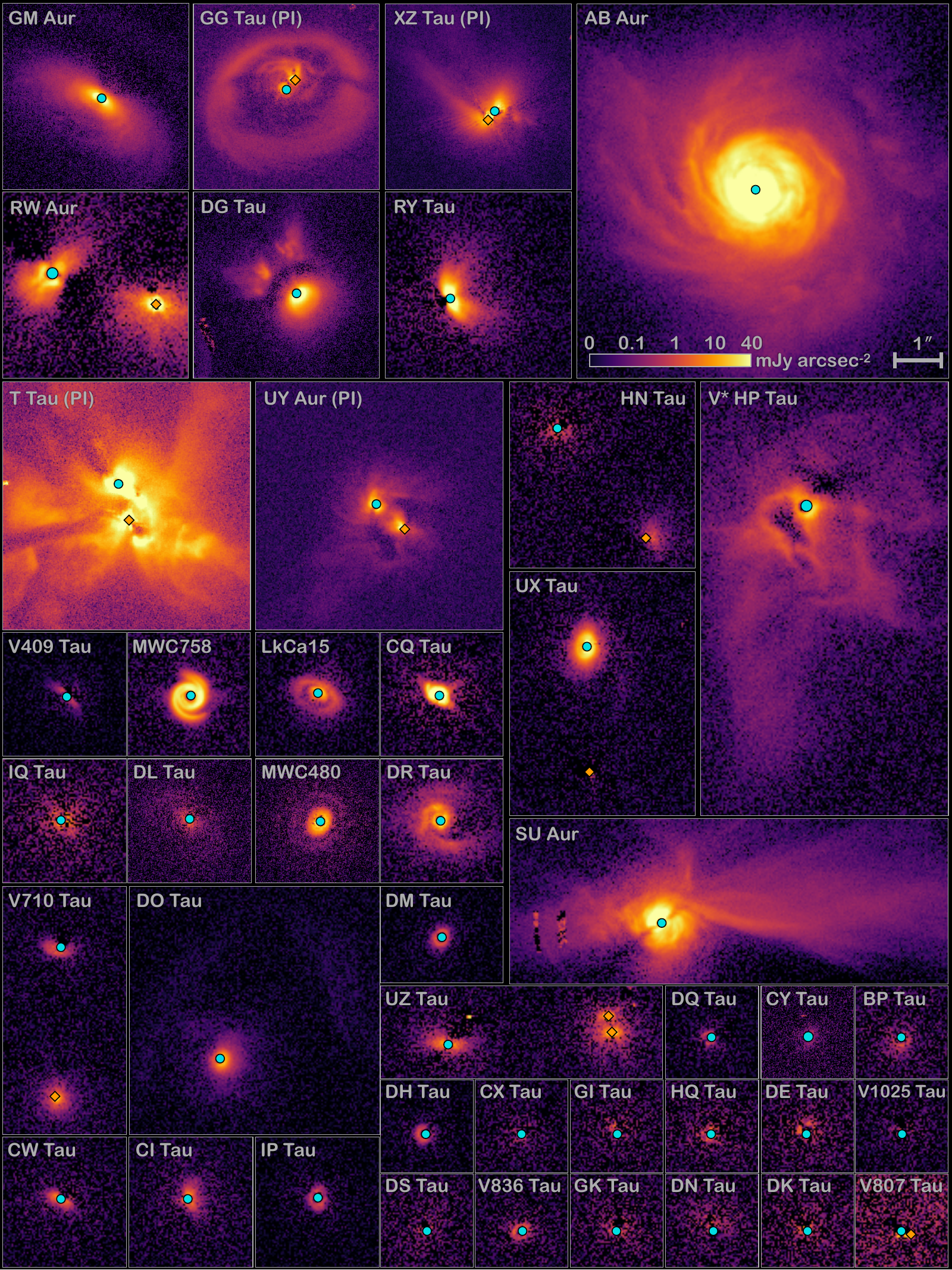} 
      \caption{Imagery of sample. The polarization map of all sources is shown with the same angular and flux scales, as indicated in the top right panel. The $Q_\phi$ parameter is shown for all sources except GG Tau, XZ Tau, T Tau, and UY Aur, for which the $PI$ parameter is shown. All primary stars masked by the coronagraph are indicated by a cyan circle, while all secondary stars are indicated by an orange diamond. Each panel is cropped to show all the significant flux detected. The only exception is T Tau, where flux is detected out to the detector edge and for which only the central region is shown. Stars are coarsely sorted by polarized light brightness.}
          \label{fig:Imagery}
  \end{figure*}

\subsection{Overview} \label{sec:overview}
The variety of brightness, flux distribution and extent shown by our large sample is appreciable from Fig.\,\ref{fig:Imagery}, where the polarized light from each source is shown in approximate decreasing order of $\alpha_{\rm pol}$. This contrast spans nearly two orders of magnitude, from \mbox{$1.6\cdot10^{-2}$} (GM Aur) to $3\cdot10^{-4}$ (the sources in the small bottom right panels of Fig.\,\ref{fig:Imagery}). Four sources (GG Tau, XZ Tau, T Tau, and UY Aur) show very strong signal in the $U_\phi$ image as well as extended negative signal in the $Q_\phi$. These are all systems where the secondary star is in close proximity to the primary (less than 1\arcsec). These companions contribute to the illumination of the circumstellar material, and therefore the observed polarized light is the result of a complex summation of the individual sources of photons \citep{Weber2023}. For these sources, we therefore show the $PI$ image, whereas for all other sources we show the $Q_\phi$ image.  

Concretely, we can divide the whole sample into three main categories: faint disks, bright disks, and targets where (much of) the scattered light originates from the surrounding environment rather than the disk. The $\alpha_{\rm pol}$ measurement and the category of each source can be found in Table \ref{table:disk_sample}. 

\subsubsection{Faint disks} \label{sec:faint}
By defining a threshold\footnote{This threshold is an approximate median of the whole distribution of $\alpha_{\rm pol}$ calculated \cite[see][]{Garufi2022b}.} of $\alpha_{\rm pol}<3\cdot10^{-3}$ , we determine the category of faint and non-detected disks. This is the largest category of the sample (27 sources corresponding to 63\% of the sample). Nonetheless, only two of these sources currently have SPHERE publications (CI Tau by \citealp{Garufi2020b} in the context of a faint disk survey and DH Tau by \citealp{vanHolstein2021} in a survey of secondary stars with possible polarimetric signal). The class of faint disks can in turn be split between sources where the disk is clearly detected and the basic disk geometry can be inferred {(see Appendix \ref{appendix:individual})} and those where the polarimetric signal is barely detected. Eleven sources belong to the former subcategory and eleven belong to the latter. Finally, five disks are formally undetected. 

The most notable characteristics of the faint disks is the absence of any obvious substructures from their image. The nearly perfect correspondence between featureless and faint disks indicate that the absence of clear substructures is merely due to our sensitivity and that elusive substructures may still be detectable (see Sect.\,\,\ref{sec:ALMA_shadowed}). There is, however, one interesting exception to this trend. MWC480 hosts one of the 12 faintest disks in the sample in terms of contrast, but a ring-like structure is clearly visible in the image (see Fig.\,\ref{fig:Imagery}). In fact, in absolute terms the disk is significantly brighter than the others. This is explained by the high stellar luminosity compared to the other faint disks (approximately 20 $\rm L_\odot$ against an average 1.2 $\rm L_\odot$). It is therefore an illustrative example of the fact that, among disks that scatter few stellar photons, substructures are more easily recovered around bright Herbig stars.   

Two other interesting cases are V409 Tau and CW Tau. As is clear from Fig.\,\ref{fig:Imagery}, these stars host an inclined disk. This information was used when determining the photospheric level for the SED (see Appendix \ref{appendix:sample}). Both stars are in fact very variable in the visible \citep[around 3 mag based on ASAS-SN curves;][]{Shappee2014}. Based on the disk geometry, we concluded that this variability may largely be due to the disk occultation, and we therefore adopted the brightest observed photometry as photospheric level. By doing so, the age calculated from the isochrones turned out to be 3.9 Myr and 3.3 Myr, respectively, which is still older than the Taurus age (possibly because the stellar flux is always partly extincted) but better than the value of $\sim$10 Myr found by adopting an average photometric value. This exercise highlights why the inferred ages of individual systems for which no resolved images are available may differ significantly from the average age of other sources in the host region. 


\subsubsection{Bright disks} \label{sec:bright}
All disks with $\alpha_{\rm pol}>3\times10^{-3}$ show evidence of substructures. The two old sources of the HD35187 association, MWC758 and CQ Tau, have disks with spirals arms \citep[see][]{Benisty2015, Hammond2022}. The disk of GG Tau is highly structured with a large cavity, shadows, spirals, and streamers \citep[see][]{Itoh2014, Keppler2020}. LkCa15 can perhaps be interpreted as a smaller version of GG Tau, exhibiting a thick ring and some inner material within the disk cavity \citep[e.g.,][]{Thalmann2016}. 

The bowl-shape of the GM Aur disk is clear from the unpublished SPHERE image (see Appendix \ref{appendix:individual}), where the disk back side and some spiral-like structures are visible \citep[see also the Subaru image by][]{Oh2016}. Spirals are also detected from the ALMA molecular emission and are probed out to more than 1000 au, suggesting a strong interaction with the cloud \citep{Huang2021}. AB Aur hosts the largest disk of the sample, and possibly of the solar neighborhood. The multiple spiral arms detected from the SPHERE image \citep{Boccaletti2020} are part of a complex, extended structure that is suggestive of a late-stage infall of material on the star \citep{Tang2012, Gupta2023}. Similarly, DR Tau shows spiral- and arc-like structures that have been associated with infall from a late encounter with a cloudlet \citep{Mesa2022}. A bright disk with evidence of external infall is also seen in SU Aur \citep{Ginski2021} where the late accretion process perturbs the disk and is responsible for spirals and shadows. In view of this, the bright disk of GM Aur, AB Aur, DR Tau, and SU Aur could also be classified in the category of ambient material. 

\subsubsection{Ambient material} \label{sec:ambient}
Some extended signal not originating from the disk surface is detected from a dozen objects. The prototype of this category is \mbox{T Tau} \citep[e.g.,][]{Kasper2020}, where a plethora of extended structures dominate in the polarimetric map and where the disks of the N, Sa, and Sb stars are hardly recognizable. The image of HP Tau shows a similar morphology. In three cases (UX Tau, UY Aur, and XZ Tau), the presence of extended arcs is to be ascribed to the binary interaction. In the prototypical example of UX Tau in particular, a bridge between the A and the C components is detected by SPHERE and by ALMA \citep{Menard2020}. Some interaction between the individual components is also revealed within the cavity of the bright disk of GG Tau \citep{Keppler2020}.

Signatures of outflow activity or, possibly, disk winds are revealed in four objects. The outflow cavity wall is visible in RY Tau \citep{Garufi2019, Valegard2022}, DO Tau \citep{Huang2022}, and RW Aur. On top of this, DO Tau also shows very extended arms probing a possible interaction with the nearby HV Tau \citep{Winter2018}. In general, vertically extended material entrained in a disk wind or natal envelope may blend with the disk signal. The net effect is the appearance of a smooth, very flared disk like those observed in DG Tau, RY Tau, and DO Tau (see Fig.\,\ref{fig:Imagery}). These objects probably reveal the first stages of the Class II phase when the optical photons from the star and the disk surface can be observed. 

\begin{figure}
  \centering
 \includegraphics[width=9cm]{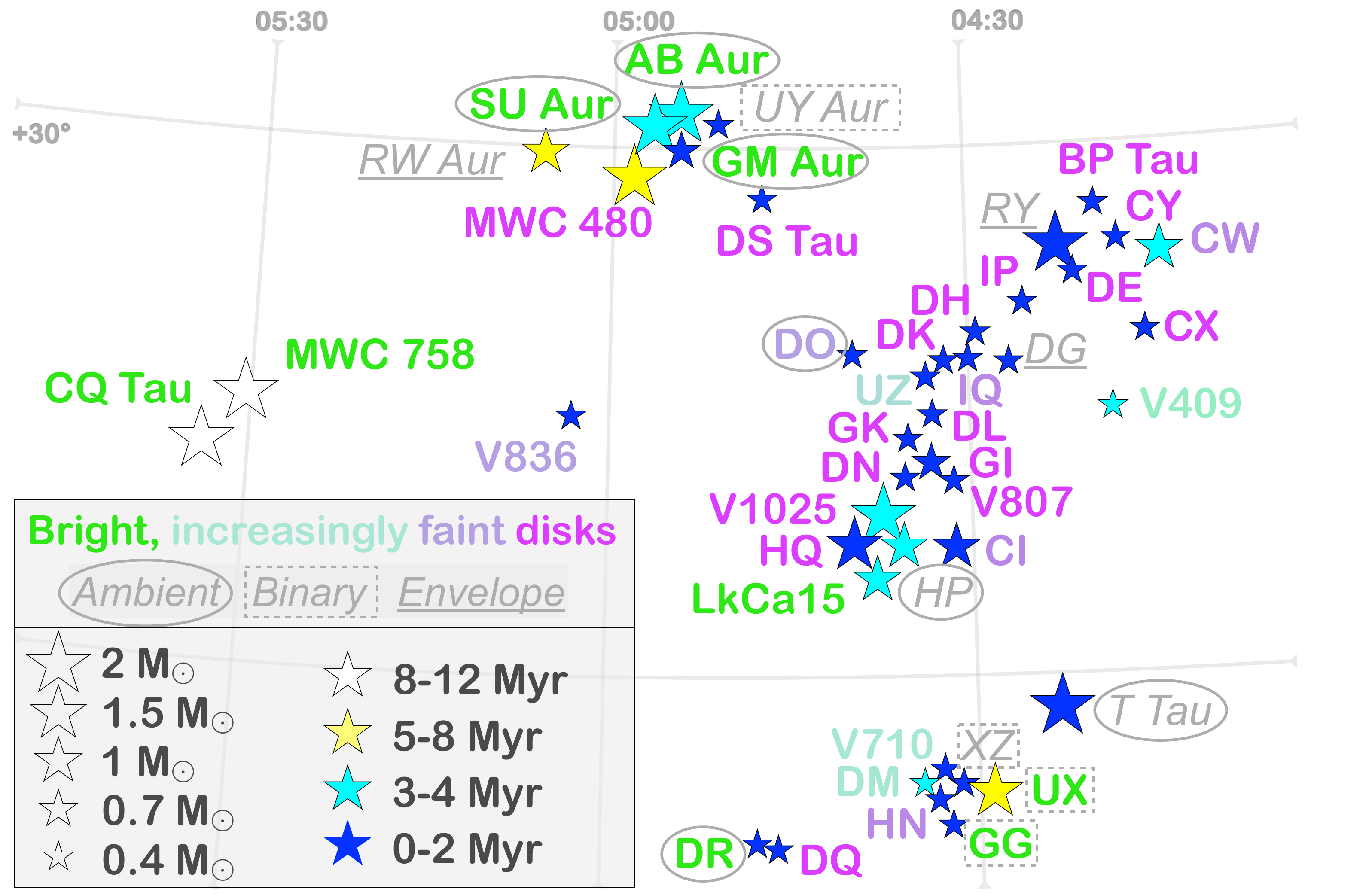} 
      \caption{Spatial distribution of stellar and disk properties. A simplified version of the map from Fig.\,\ref{fig:Taurus_map} is shown to highlight the distribution of bright disks (green), increasingly faint disks (fading to purple), of ambient material (rounded boxes), of binary interaction (rectangular boxes) and of envelope emission (underline), as well of stellar mass (symbol size), and age (symbol color).}
          \label{fig:properties_map}
  \end{figure}

\subsection{Statistics and spatial distribution} \label{sec:statistics}
In Fig.\,\ref{fig:properties_map}, we show the distribution of the various categories across the Taurus region. Evidently, bright and isolated disks in the sample are rare. In fact, there are only three stars hosting a disk with substructures and no evidence of interaction with the environment. Since two of these, CQ Tau and MWC758, are formally not in Taurus, this census is limited to LkCa15. However, if we also include bright disks around binary stars in mutual interaction (two, see Sect.\,\ref{sec:ambient}) and around single stars in possible interaction with the environment (four, see Sect.\,\ref{sec:bright}), this count increases to seven (16\% of the sample). Conversely, targets with faint disks and no sign of interaction are 26 (60\%), plus two (summing up to 65\%) in interaction with the environment. Regardless of the disk brightness, six targets (14\%) show some interaction with the environment, while four sources (9\%) show an envelope and detectable outflow activity. Excluding the spectroscopic binary DQ Tau, there are 12 multiple systems in the sample. Of these, five (42\%) show some interaction with the individual components.

The map in Fig.\,\ref{fig:properties_map} also reveals a hint of segregation between the central portion of Taurus and the peripheral regions. In the central portion of Taurus where our most represented groups L1495, L1524, and L1536 lie, mostly isolated and faint disks are found. {This segregation cannot be driven by the interstellar extinction since this affects equally the stellar and disk signal considered in the contrast calculation (see Sect.\,\ref{sec:Observations}). The absolute effect on the observed disk brightness is also minor since in the H band the extinction $A_{\rm H}$ is only one-fifth of the $A_{\rm V}$, and the $A_{\rm V}$ of most of our sample is lower than 2 mag.} The only bright disk in this portion of Taurus that comprises 24 sources is LkCa15 (in the L1536 group at the southern edge of this central region). Interestingly, LkCa15 is also the only star in this main region that appears significantly older (6-9 Myr) from our isochrone calculation. HP Tau and DO Tau are the only sources of the central region clearly interacting with the environment, and both lie on the edges of the region, to the south and to east, respectively.

The northern L1517 shows a very different population of objects from our sample, hosting as many as four bright disks out of seven members, with three stars, GM Aur, AB Aur, and SU Aur, that are interacting with the environment (and two of which, AB Aur and SU Aur, are in relative proximity). Similarly, the southern portion of Taurus with the L1551, L1558, and T Tau groups have a high incidence of bright disks (with only 2 out of 9 very faint disks), and of ambient material (5 out of 9).

\subsection{Correlation with stellar and disk properties} \label{sec:trends}
\begin{figure}
  \centering
 \includegraphics[width=8cm]{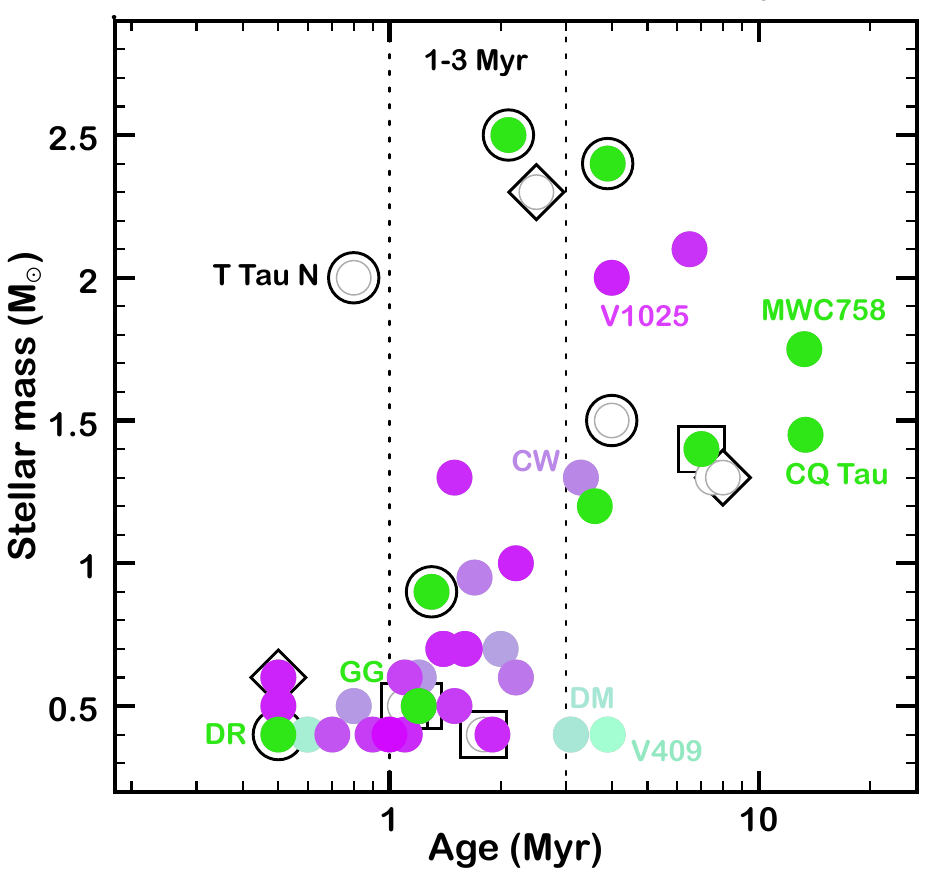}
 \includegraphics[width=8cm]{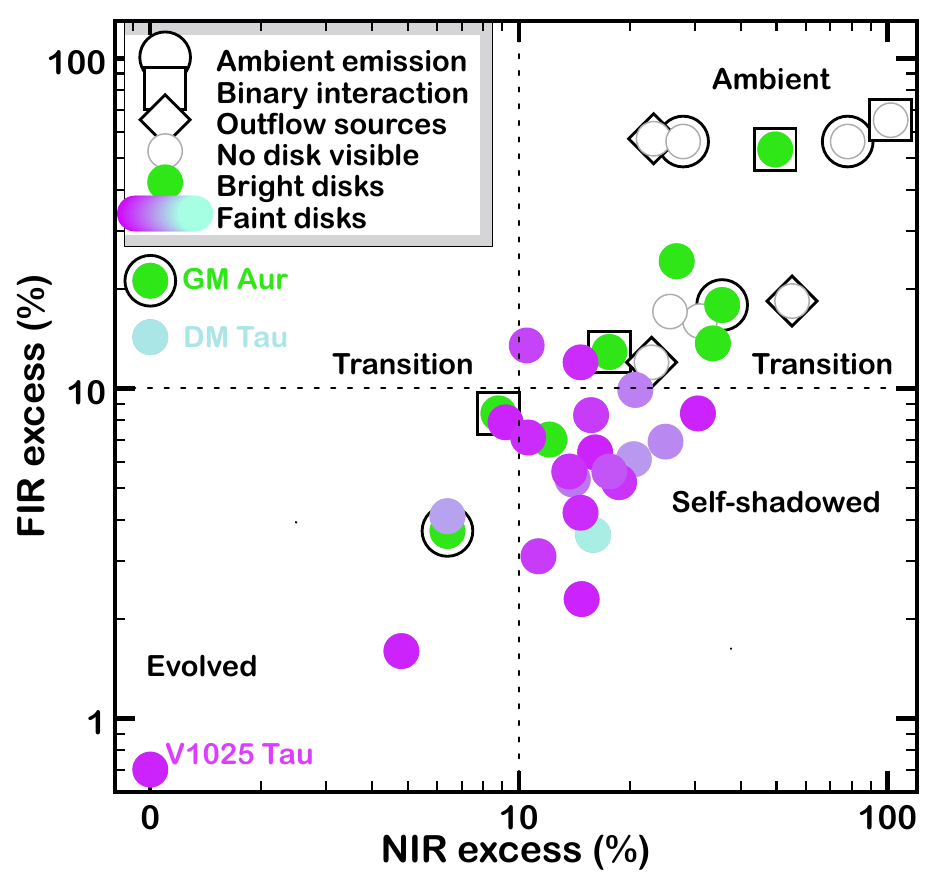}
 \includegraphics[width=8cm]{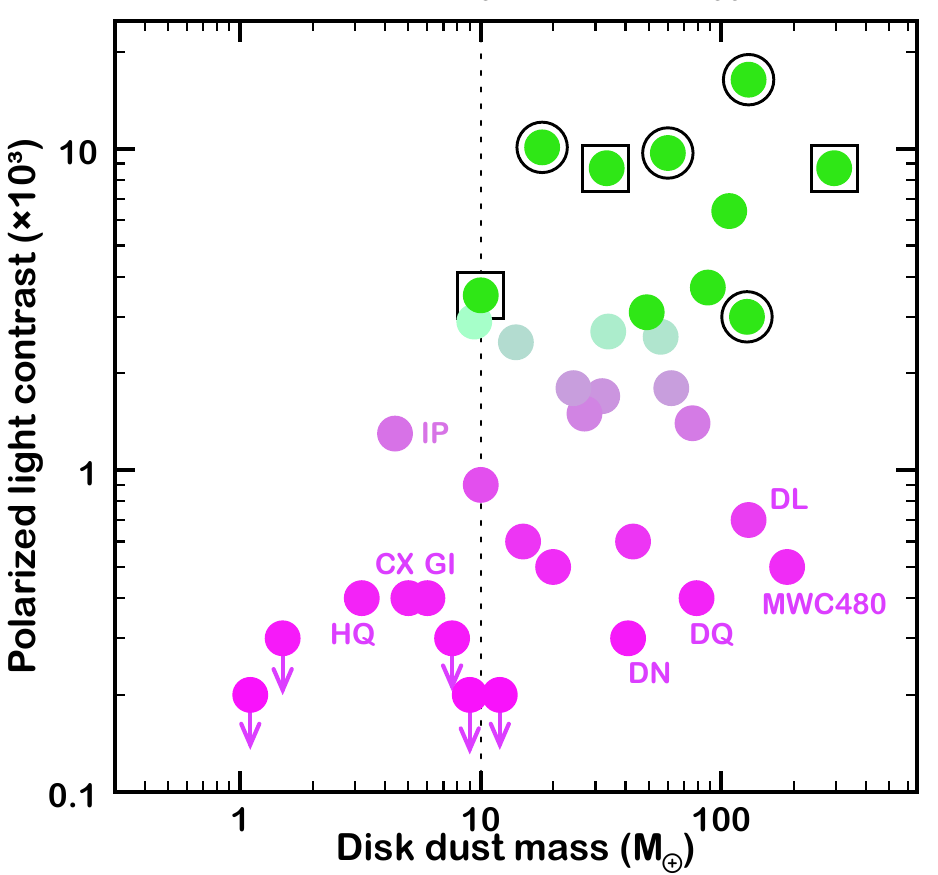}
      \caption{Disk categories across stellar and disk properties. Top: Stellar mass versus age. The dashed lines delimit the Taurus age \citep[1--3 Myr,][]{Luhman2023}. Middle: FIR versus NIR excess, shown in fraction of the stellar luminosity. The dashed lines indicate the ideal separation between low and high excesses. Bottom: Polarized light contrast versus disk dust mass. The dashed line indicates the detection limit from \citet{Garufi2022b}.}
          \label{fig:trends}
  \end{figure}

The three categories described in Sect.\,\ref{sec:overview} and the disk brightness in scattered light (see Sect.\,\ref{sec:Observations}) are compared with stellar and disk properties in Fig.\,\ref{fig:trends}, and the results are discussed in this section.

\subsubsection{Stellar age and mass} \label{sec:trends_stellar}
Similarly to what is seen in Fig.\,\ref{fig:properties_map}, any relation between the disk brightness and the stellar properties can be appreciated from Fig.\,\ref{fig:trends} (top panel).  From the diagram, it is clear that the majority of stars show an age in line with Taurus \citep[$<$3 Myr; e.g.,][]{Luhman2023}. On the one hand, most of the young stars are of low mass and host a faint disk, with the only bright disk being found in GG Tau (a multiple system) and DR Tau (in strong interaction with the environment). On the other hand, a large fraction of the old stars show a bright disk. This dichotomy may suggest that the old age calculated for these sources is real.

Beside MWC758 and CQ Tau (associated with an older group; see Sect.\,\ref{sec:sample}), we find ten sources older than 3 Myr. Two of these, CW Tau and V409 Tau, host an inclined disk that may occult some stellar flux yielding a biased age (see Sect.\,\ref{sec:overview}). A similar conclusion may in principle be drawn for RW Aur and HP Tau with strong ambient emission. Two of the remaining old stars, DM Tau and V1025 Tau are interesting cases that are discussed below.

It must also be noted that the sample does not contain any young, intermediate-mass stars, with the notable exception of T Tau. Also, among the few intermediate-mass stars we find most (4 out of 5) of the sources evidently interacting with the environment.

\subsubsection{Infrared excess} \label{sec:trends_excess}
\citet{Garufi2022b} adopted a threshold of 10\% to separate high and low IR excesses resulting in a NIR-FIR diagram divided into four quadrants. These quadrants host stars with different disk types. As labeled in Fig.\,\ref{fig:trends} (middle panel), these are transition disks (with both high NIR and low NIR), self-shadowed disks, and evolved disks. Almost all the faint disks of the sample sit in the self-shadowed quadrant. This reflects the geometry where a robust NIR excess (10--20\%) originates from the inner disk, while a small FIR excess (<10\%) is reprocessed by the outer disk in penumbra \citep[see][]{Garufi2020b}. Conversely, mostly bright disks and sources with ambient emission sit in the upper right panel.

Nonetheless, a major difference from the large sample by \citet{Garufi2022b} arises, that is the very low number of sources with low NIR excess and high FIR excess (top left quadrant). This is by itself an interesting finding as it points to the substantial absence of massive disks with depleted inner regions. The only two objects of the whole quadrant are GM Aur and DM Tau. Instead, to the bottom left we find V1025 Tau with nearly null IR excess (pointing to an evolved Class III target). Interestingly, DM Tau and V1025 Tau are the two old faint disks mentioned in Sect.\,\ref{sec:trends_stellar}, suggesting a causality between the two findings, which is discussed in Sect.\,\ref{Discussion}.

\subsubsection{Disk mass}
As is clear from Fig.\,\ref{fig:trends} (bottom panel), there is no tight trend between the disk brightness in scattered light and the dust disk mass. There are many bright and many faint SPHERE images of disks with dust masses higher than 10 M$_\oplus$. However, disks less massive than 10 M$_\oplus$ are barely detected, in agreement with the findings of \citet{Garufi2022b}. In this regard, the detections around GI Tau, CX Tau, HQ Tau, and IP Tau (all with less than 6 M$_\oplus$) are notable as these are among the least massive disks ever resolved in polarized light. The disk of IP Tau in particular is also relatively bright given the low mass (see also Sect.\,\ref{sec:ALMA_cavity}).

This diagram also provides a qualitative assessment of why a disk is undetected. On the one hand, disks less massive than 5--10 M$_\oplus$ are probably small and would therefore scatter most photons at separations not achievable by the current telescopes. On the other hand, disks more massive than 10 M$_\oplus$ are most likely extended, and their non-detection is due to the self-shadowed geometry of the disk (see Sect.\,\ref{sec:ALMA_shadowed}).

\section{Results from the comparison with ALMA} \label{sec:ALMA}
At the time of writing, all the sources of this work, except V1025 Tau, have been observed with ALMA. More than half of the sample has been observed within the Taurus snapshot survey described by \citet{Long2018b, Long2019} and \citet{Manara2019}. Their sample was built excluding sources with ALMA high-resolution data available at the time. Given that bright, well-known sources were first targeted by the ALMA users, it is not surprising that most of the \citet{Long2019} targets are faint disks in scattered light. In this section, we give a brief review on the continuum ALMA images deferring any comparison with molecular emission to a future work. In particular, we focus on a few interesting cases for which no extensive studies have been reported yet.

\subsection{Census of the ALMA sample}
Out of the 42 disks with high-resolution ALMA data, nine show a large inner cavity. It is well established that the presence of a large cavity devoid of dust is the primary condition for a strong scattered light signal \citep[see e.g.,][]{Garufi2017}, and indeed seven of these nine disks with a cavity are bright. These seven bright disks have been studied in previous ALMA work \citep{Tang2017, Boehler2018, UbeiraGabellini2019, Facchini2020, Huang2020, Francis2020, Wolfer2021}, and present a geometry where NIR scattered light is detected within the ALMA cavity, revealing spirals \citep[in CQ Tau, MWC758, GG Tau, AB Aur, and GM Aur,][and Appendix \ref{appendix:individual}, respectively]{Hammond2022, Benisty2015, Keppler2020, Boccaletti2020} or some degree of asymmetry \citep[in LkCa15 and UX Tau]{Thalmann2016, Menard2020}. However, the two faint disks with an ALMA cavity (IP Tau and DM Tau) are also interesting cases that are described in Sect.\,\ref{sec:ALMA_cavity}.

Out of the remaining 33 disks with no large cavity, nine show obvious evidence of substructures in the currently available ALMA data. Seven of these nine are faint disks that are described in Sect.\,\ref{sec:ALMA_shadowed}. The remaining two (DG Tau and RY Tau) are the prototypical example of partly embedded disks (see Sect.\,\ref{sec:ambient}) for which the NIR images only yield limited information on the disk surface. They are showcases of the early formation of substructures in disks \citep[as in the pioneering example of HL Tau,][]{ALMA2015}.

As many as 24 ALMA disks can be coarsely considered featureless for the time being, although the diverse angular resolution and sensitivity of the various datasets nullify any quantitative assessment of the class of objects. Future improved ALMA observations or reanalysis of the current sample \citep[see e.g.,][]{Jennings2022, Zhang2023} may reveal further substructures. All these disks are either faint in scattered light or dominated by ambient emission. Eight of these disks are part of a multiple system as described in Sect.\,\ref{sec:ALMA_systems}. Some additional information about the individual cases are given in Appendix \ref{appendix:individual}.

\begin{figure}
  \centering
 \includegraphics[width=9cm]{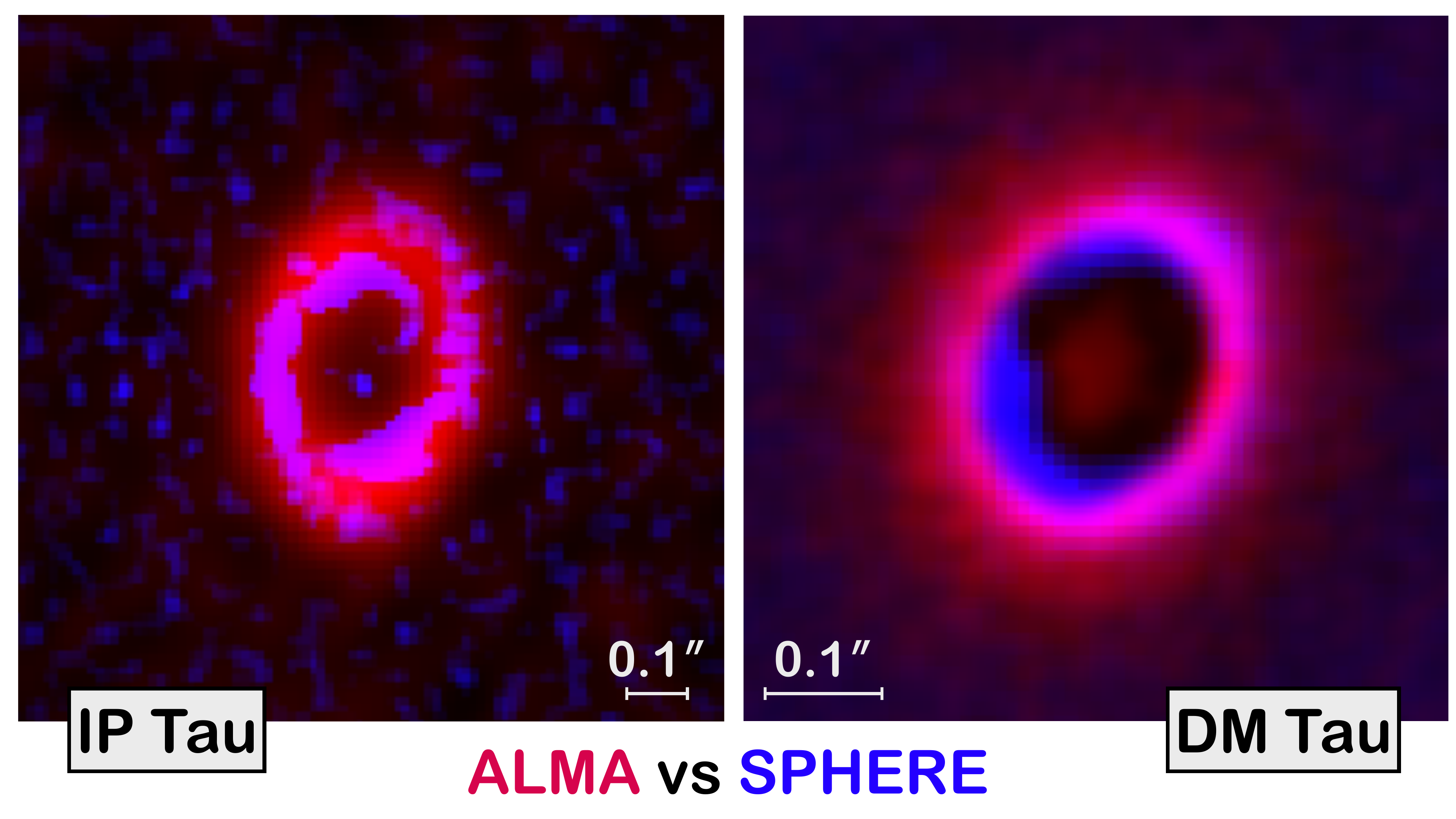}
      \caption{Disks with an ALMA cavity that are faint in scattered light. The ALMA image (in red) of IP Tau and DM Tau is compared to the high-pass filtered SPHERE image (in blue). The innermost region of the original SPHERE images is masked by the coronagraph.}
          \label{fig:faint_cavity}
  \end{figure}

\subsection{The peculiar cases of IP Tau and DM Tau} \label{sec:ALMA_cavity}
The two disks with a cavity from the ALMA continuum that appear faint from our SPHERE observations are IP Tau and DM Tau. They are both, per se, interesting cases, as described in Sect.\,\ref{sec:trends}. As is clear from Fig.\,\ref{fig:trends}, DM Tau is the only low-mass star that looks older than 3 Myr\footnote{Excluding V409 Tau, for which the disk inclination may bias the age estimate (see Sect.\,\ref{sec:stellar_properties}).} and is among the few Class II sources ever observed with null NIR excess that is not particularly bright in scattered light. IP Tau is the brightest disk in the sample with less than 10 M$_\oplus$ in dust mass (see Fig.\,\ref{fig:trends}), and is probably among the brightest ever observed of such a type.

In Fig.\,\ref{fig:faint_cavity}, we compare the SPHERE image of IP Tau and DM Tau with the ALMA continuum by \citet{Long2019} and \citet{Hashimoto2021}. We show a high-pass-filtered version of the SPHERE image obtained by convolving the image with a Gaussian profile and then removing the convolved image from the original image. This is a common technique to highlight elusive substructures. This exercise clearly reveals the presence of spirals in the disk of IP Tau that are co-spatial with the inner edge of the ALMA disk. In DM Tau, a clear asymmetry is visible to the SE where the inner edge of the NIR emission seems to penetrate deeper in the ALMA cavity with respect to the symmetrical portion around the minor axis, to NW. In principle, this asymmetry may also point to an unresolved feature such as a spiral. These findings are in line with the recurrent presence of such substructures in disks with a cavity described above \citep[see also][]{vanderMarel2021}, and they are further discussed in Sect.\,\ref{Discussion}.  

\begin{figure}
  \centering
 \includegraphics[width=8.3cm]{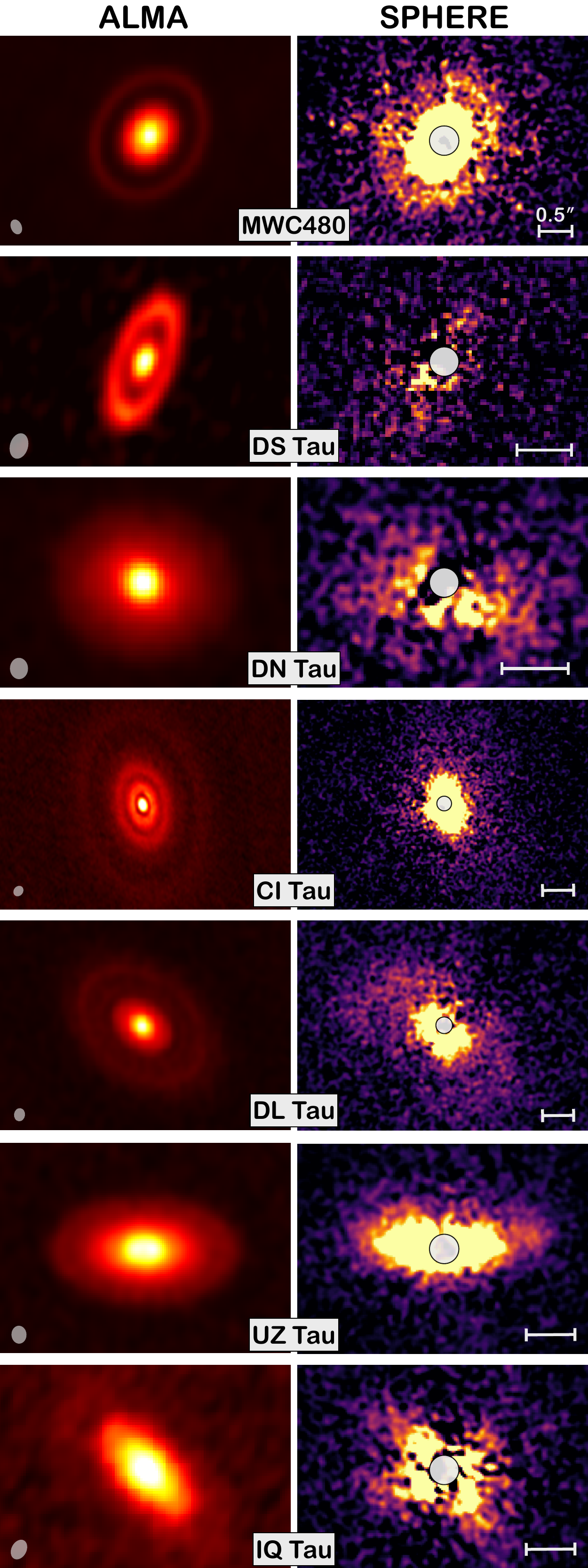}
      \caption{Self-shadowed disks in sample. The ALMA image of those disks that show evidence of substructures is placed beside an intentionally saturated SPHERE image obtained with a light smoothing. The ALMA beam is shown in the bottom right of each image, the physical scale (0.5\arcsec) common to ALMA and SPHERE relative panels can be seen in the bottom left. All ALMA images are from \citet{Long2019}, except CI Tau, which is from \citet{Rosotti2021}.}
          \label{fig:shadowed}
  \end{figure}

\subsection{The cases of the self-shadowed disks} \label{sec:ALMA_shadowed}
In Fig.\,\ref{fig:shadowed}, we compare ALMA and SPHERE images of the seven faint disks in our sample that are classified as substructured disks by \citet{Long2019}. Generally speaking, the comparison reveals that only a fraction of the disk is probed by the SPHERE observations. This is particularly true for CI Tau \citep[see also][]{Garufi2022b} and DN Tau. This configuration is the most direct proof that these disks are self-shadowed, like formulated by \citet{Dullemond2004a} and shown observationally by \citet{Garufi2022b}.

Other examples in Fig.\,\ref{fig:shadowed}, such as UZ Tau and IQ Tau, are similar. The SPHERE signal extent is comparable to ALMA, but it is reasonable to assume that the gas disk extent is larger than the ALMA continuum and is therefore also not probed by our observations. In the case of DL Tau, as in MWC480, some ring-like structures do become visible after light smoothing of the image and aided by the comparison with ALMA. Finally, in the case of DS Tau the comparison with ALMA allows us to determine the presence of a weak $Q_\phi$ signal consistent with disk scattered light, even though this target is formally treated as a non-detection in this work (see Appendix \ref{appendix:individual}). 

\begin{figure*}
  \centering
  \includegraphics[width=17cm]{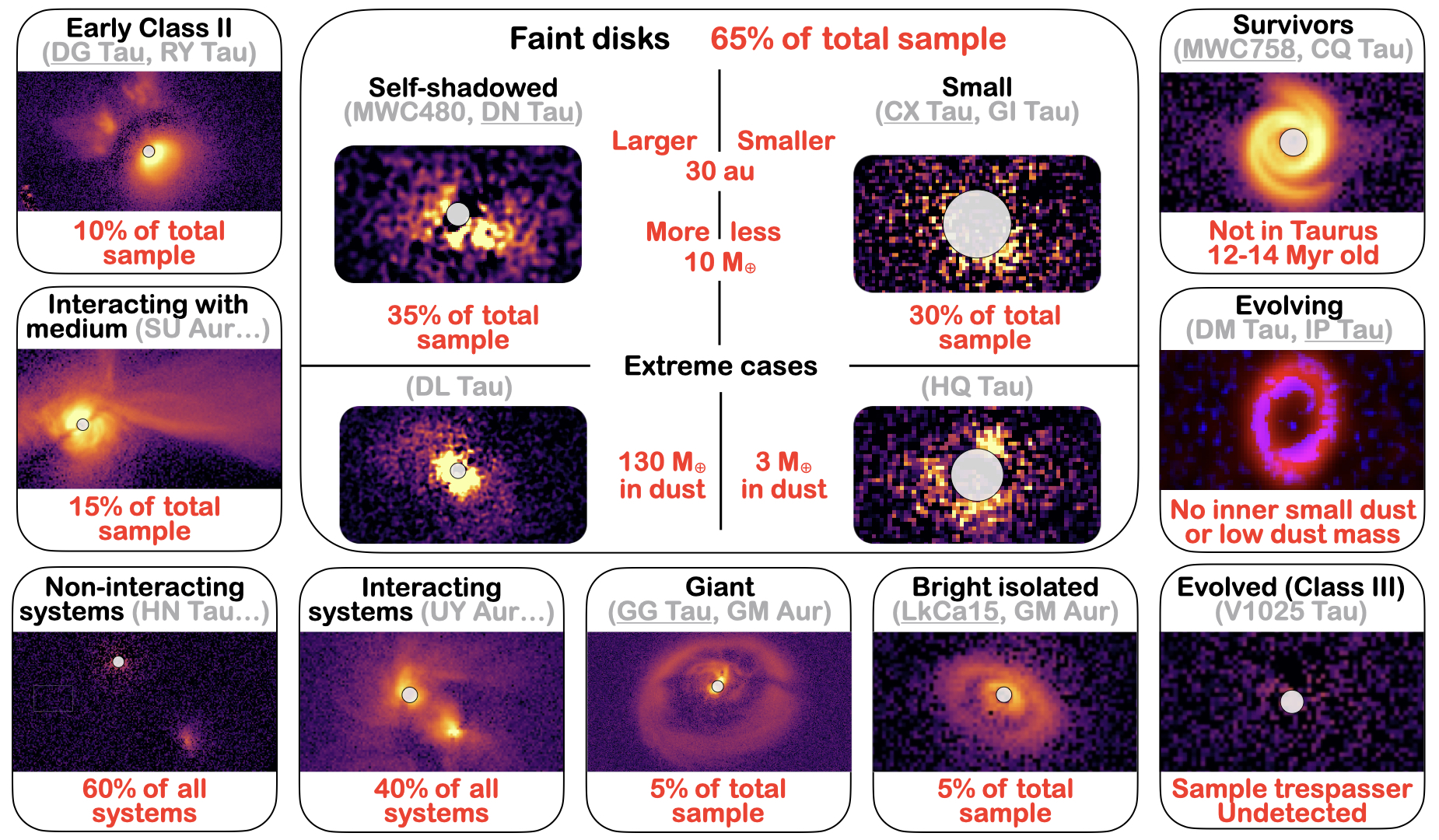}
      \caption{Visual synopsis of sample. The demography of the SPHERE sample studied in this work is summarized with illustrative examples. The main central box includes the prototypical object of the sample, that is, a faint, isolated disk. All the peculiar cases that do not belong to this category are depicted in the smaller boxes. The name of the source shown is underlined.}
          \label{fig:sketch}
\end{figure*}

\subsection{The cases of multiple systems} \label{sec:ALMA_systems}
Only three of our disks in multiple systems are extended by more than 0.2\arcsec\ from the ALMA continuum (UX Tau A, UZ Tau, and V710 Tau). In each of these three systems, the separation between the individual components is large (>2.7\arcsec). These three disks are clearly resolved with SPHERE. Furthermore, a disk is also resolved around UX Tau C \citep[see][]{Menard2020} and UZ Tau B and C (see Fig.\,\ref{fig:Imagery}), where two disks with the same position angle of the primary disk \citep[in line with the ALMA image by][]{Manara2019} are resolved out to 0.4\arcsec\ and 0.2\arcsec, respectively.

In the remaining eight cases, ALMA constrains a small disk (<0.15\arcsec) with a steep decay at the outer radius that is indicative of tidal truncation \citep{Manara2019}. In only one of these cases, DH Tau A, the disk is clearly resolved by SPHERE \citep[beside the detection of the disk around DH Tau B by][]{vanHolstein2021}, while in one case, RW Aur, some disk emission is embedded with some ambient material (see Sect.\,\ref{sec:ambient}). In all other cases the disk is not detected either because it is hidden by the interacting material (XZ Tau, UY Aur, T Tau) or because no significant polarimetric signal is visible (HN Tau, V807 Tau, DK Tau).

\section{Discussion} \label{Discussion}
This work contributes to the general effort of the planet-forming disk community to characterize less and less exceptional objects and to study the disk demography as a whole. Thanks to the work put in place by many authors, a very large fraction of the observable Taurus population has now been observed. In this work, we show that the current sample is relatively complete for stars with masses above 0.4 M$_\odot$. It does, however, also urge the community to develop NIR facilities that can probe optically fainter stars as this means covering more embedded, lower-mass, and further away objects.

One of the key results of this work is that the prototypical disk in Taurus (1--3 Myr) appears faint in scattered light. This is partly due to the small physical extent of some disks and partly to the low incidence of an inner cavity in these disks that would allow the efficient illumination of the outer disk regions. In fact, the partition between disks with and without a large cavity from ALMA images naturally demarcates our boundary between bright and faint disks from the SPHERE images. A large fraction of these faint disks will never form an inner cavity, as suggested by complete Spitzer survey and ALMA demographical studies \citep[respectively]{Williams2011, Cieza2019, Cieza2021}. Nonetheless, the fraction of bright disks from surveys of Class II sources only is still expected to increase with time because disks that do not form a cavity will turn into evolved Class III sources, leaving only disks with the cavity in the sample.

In this work, we also find a possible spatial segregation for different disk types within the whole region, with the central portion of Taurus mainly hosting faint, isolated disks and with the Taurus peripheries hosting a higher number of interacting, bright disks. It is tempting to draw an evolutionary history for the region that is, however, not captured yet by the estimated ages of the individual Taurus groups. It could instead be linked to the higher stellar mass of some peripheral group members (see, e.g., SU Aur, AB Aur or T Tau and UX Tau) for which a higher incidence of interaction with the environment is found in this work or a faster disk evolution is reasonably expected.

Understanding the disk demography from a uniform sample like this in Taurus is also beneficial for studying the outliers and the extreme cases. A visual synopsis of the peculiar objects encountered in this work is given in Fig.\,\ref{fig:sketch}. The main box is reserved to the large portion of faint disks in the sample. The disk extent constrained from the ALMA continuum or the dust mass measured from the millimeter photometry aid the interpretation of the actual disk geometry. On the one hand extended, self-shadowed disks are well represented in the Taurus sample with the most extreme cases being MWC480 and DL Tau (that are very massive but are relatively easily detected by SPHERE) or DN Tau and DQ Tau (that are barely detected by SPHERE but are only mildly massive). On the other hand, small disks that are less than 10 M$_\oplus$ in dust mass are hardly detectable by SPHERE. The resolved detection of as many as four disks less massive than that (HQ, CX, GI, and IP Tau) is an important step toward the characterization of less exceptional objects.

On the opposite end of the dust mass distribution we find the well-known giant disks such as those of GG Tau and GM Aur. This work highlights their low incidence since the sample bias is alleviated, and it shows that there is a portion no larger than 20\%. This portion is reduced to just 5\% if we only consider isolated, bright disks and exclude the two old members of a nearby association, MWC758 and CQ Tau. It is tempting and possibly reasonable to consider these disks the only survivors of an ancient disk population with an estimated age of 18 Myr. In any case, the fraction of bright disks could be even smaller when the population of lower mass stars not observed by SPHERE is considered. 

It is increasingly clear to the community that planet-forming disks are not isolated structures. The incidence of targets that are interacting with the environment is another pivotal element in disk evolution studies that can only be constrained by looking at an unbiased sample. The 30\% portion that is measured in this work could, however, be smaller when lower mass stars are considered. Indeed, we find that the majority of stars interacting with their environment are of intermediate mass, although this may be partly be due to the higher luminosity of these stars that can reveal ambient material on a larger volume. In addition, a large portion (40\%) of stars in multiple systems show mutual interactions. 

Finally, it is illustrative to look at the beginning and end stages of evolution within our sample of Class II. The morphology observed in objects such as DG Tau and RY Tau suggests that we are seeing young Class II objects where optical photons from the star and the disk can reach the observer, although still blended with the contribution from natal envelope and outflows. As the other extreme, we propose that two objects of the sample, DM Tau and IP Tau, represent the final stage of the Class II phase. Both are relatively faint in scattered light despite the presence of a large ALMA cavity. DM Tau shows no NIR excess from the SED despite an inner disk component that is detected by ALMA \citep{Hashimoto2021}, pointing to a population of large grains that are only in the inner region. It is also the only low-mass star in the sample that appears to be older than 3 Myr. Given the small uncertainties relative to this measurement (no extinction, precise distance), we consider this estimate correct. IP Tau hosts a disk with a very small amount of dust, but, in this regard, it is still relatively bright in scattered light. The presence of spirals detected in the SPHERE image reveals complex dynamical processes as in more massive, brighter transition disks \citep{vanderMarel2021}, but the low disk mass compared to the extent of the scattered light emission suggests that this object is transitioning to the Class III stage, and this is the only such case in our sample of V1025 Tau.

\section{Summary} \label{Summary}
In this work, we examined the demographics of planet-forming disks in Taurus through the full SPHERE dataset available. This sums up to 43 sources with NIR polarimetric images (31 unpublished) that probe the scattered light from any circumstellar material. Our main findings are the following.

\begin{itemize}
    \item The available sample covers approximately one-fifth of the total Class II population in Taurus and approximately a half of those that can be observed with SPHERE (three quarters of those observable in optimal conditions). All in all, the sample is complete above 0.4 M$_\odot,$ but it is totally blind to lower mass stars.
    
    \item Thirteen of the 43 targets are multiple systems. The individual components in the sample include ten intermediate-mass, nine solar-mass, and 36 low-mass stars.

    \item The majority of the sample (63\%) is made up of faint disks with no prior study from the literature. No obvious substructures are visible in these disks at the current sensitivity and resolution limits. In a few cases, rings that are clear from ALMA are recovered in the SPHERE images (e.g., MWC480, DL Tau).

    \item Disks that are less massive than 10 M$_\oplus$ are rarely detected by SPHERE because of their limited size. However, several massive disks are also barely detectable (e.g., DN Tau). These are the extended self-shadowed disks that have not (yet) formed a cavity. Their high occurrence in Taurus is likely due to their young age.  

    \item Ambient material is visible in 30\% of the sample, with a mild prevalence in intermediate-mass stars. This class includes disks in interaction with the environment (e.g., SU Aur), binaries in mutual interaction (e.g., UY Aur), and stars with evidence of natal envelope and outflow activities (e.g., RY Tau).  

    \item About 40\% of the binary systems show some evidence of interaction between the individual components. The disks are rarely detected in these systems (and never in the case of close companions). This can be reconciled with small disk sizes due to truncation proposed from previous ALMA observations.

    \item Bright, isolated disks as in, for example, LkCa15 are rare (5\%). Bright disks in general (including those in interaction with environment and companions) add up to 23\%. Only two of these disks, GM Aur and DM Tau, are significantly depleted of NIR excess from the SED. This incidence is lower than in previous study with diversified samples and points to an early evolutionary stage of Taurus sources.

    \item The presence of a disk cavity from ALMA naturally intercepts all and only bright disks in the NIR. Two interesting exceptions are the relatively faint disks of IP Tau and DM Tau, showing mild spirals and asymmetries, as the other disks with an ALMA cavity. The low dust mass of the former and the null NIR excess and old age of the latter suggest that these disks are transitioning to an evolved stage.

    \item A hint of spatial segregation for disk types is visible within the Taurus region. The central portion of the region primarily hosts faint, isolated disks, while a higher occurrence of bright, interacting disks are found in the Taurus periphery.

    \item Some stars with an inclined disk appear artificially older (e.g., V409 Tau) because of occultation of the stellar photosphere. Apart from these, it is clear that brighter disks are found around older stars. This indicates that the ages derived from isochrones are probably reasonably accurate and point to an expected evolutionary path where mostly only disks with a large cavity survive more than 3 Myr. 
    
\end{itemize}

In summary, several findings indicate an early evolutionary stage for the overall Taurus population, although some individual objects may represent advanced or exceptional evolutionary stages and some others may reveal intermediate phases of the disk lifetime. Together with the parallel works on Chamaeleon (Ginski et al.) and Orion (Valegard et al.), this is a first effort to characterize the demographics of planet-forming disks of an individual region from NIR images. Future work includes the characterization of additional star-forming regions as well as the direct comparison of the results found in different regions to determine the evolution of planet-forming disks from the largest sample currently possible.

\begin{acknowledgements}
      We thank the referee for the useful comments as well as F. Long, G. Rosotti, J. Hashimoto, and N. van der Marel for sharing their ALMA data. SPHERE is an instrument designed and built by a consortium consisting of IPAG (Grenoble, France), MPIA (Heidelberg, Germany), LAM (Marseille, France), LESIA (Paris, France), Laboratoire Lagrange (Nice, France), INAF Osservatorio di Padova (Italy), Observatoire de Gen\`{e}ve (Switzerland), ETH Zurich (Switzerland), NOVA (Netherlands), ONERA (France) and ASTRON (Netherlands) in collaboration with ESO. SPHERE was funded by ESO, with additional contributions from CNRS (France), MPIA (Germany), INAF (Italy), FINES (Switzerland) and NOVA (Netherlands). SPHERE also received funding from the European Commission Sixth and Seventh Framework Programmes as part of the Optical Infrared Coordination Network for Astronomy (OPTICON) under grant number RII3-Ct-2004-001566 for FP6 (2004–2008), grant number 226604 for FP7 (2009–2012) and grant number 312430 for FP7 (2013–2016). This research has made use of the VizieR catalogue access tool, CDS, Strasbourg, France (DOI: 10.26093/cds/vizier). The original description of the VizieR service was published in 2000, A\&AS 143, 23. This work was supported by the PRIN-INAF 2019 Planetary Systems At Early Ages (PLATEA) and by the Large Grant INAF 2022 YODA (YSOs Outflows, Disks and Accretion: towards a global framework for the evolution of planet forming systems). We also acknowledge financial support from the Programme National de Plan\`{e}tologie (PNP) and the Programme National de Physique Stellaire (PNPS) of CNRS-INSU in France and from ANID -- Millennium Science Initiative Program -- Center Code NCN2021\_080. This project has received funding from the European Research Council (ERC) under the European Union's Horizon 2020 research and innovation programme (PROTOPLANETS, grant agreement No. 101002188). C.F.M. is funded by the European Union (ERC, WANDA, 101039452). S.P. acknowledges support from FONDECYT grant 1231663. P.W. acknowledges support from FONDECYT grant 3220399. S.F. is funded by the European Union under the European Union's Horizon Europe Research \& Innovation Programme 101076613 (UNVEIL). Views and opinions expressed are however those of the author(s) only and do not necessarily reflect those of the European Union or the European Research Council. Neither the European Union nor the granting authority can be held responsible for them. This project has received funding from the European Research Council (ERC) under the European Union's Horizon Europe research and innovation program (grant agreement No. 101053020, project Dust2Planets).
\end{acknowledgements}

%
%

\bibliographystyle{aa} 
\bibliography{Reference} 

\begin{appendix} 

\section{Spatial distribution of the sample} \label{appendix:distribution}
\begin{figure}[h]
  \centering
 \includegraphics[width=9cm]{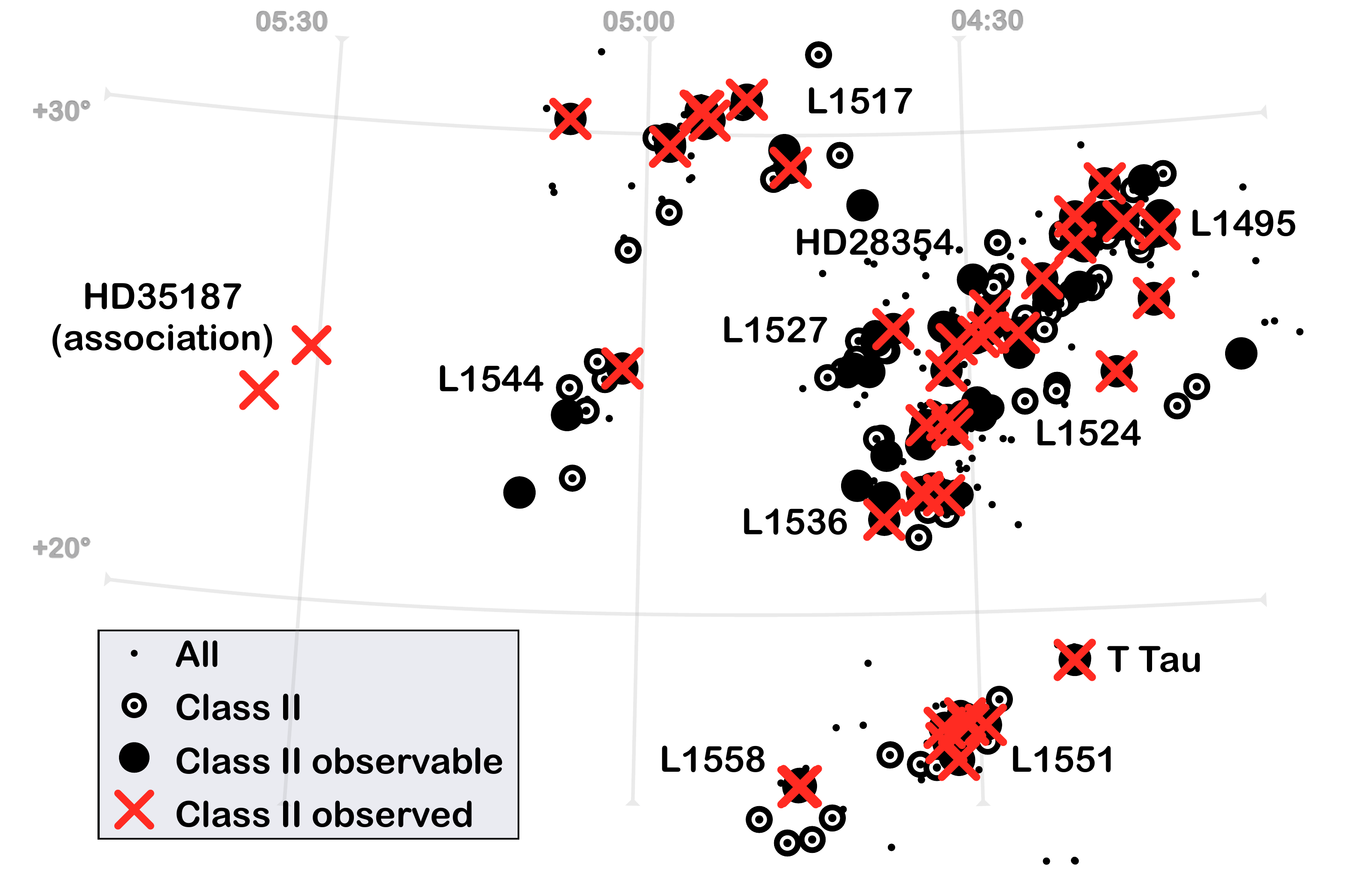} 
      \caption{Sample of this work compared with the whole Taurus. The spatial distribution of all Taurus sources from \citet{Luhman2023} is compared with the that of Class II sources, of Class II sources that are observable by SPHERE, and of those observed. The groups with at least one member in this work are labeled \citep[see][for the complete list]{Luhman2023}.}.
          \label{fig:selection_map}
  \end{figure}

The spatial distribution of the entire Taurus sample by \citet{Luhman2023} is shown in Fig.\,\ref{fig:selection_map}. The sample of this work is also shown in the map together with the Class II sources that are possibly observable by SPHERE. The names of the groups with at least one member in this work are also shown. 

\section{SEDs and properties of the sample} \label{appendix:sample}
\begin{figure*}[h]
  \centering
 \includegraphics[width=16.5cm]{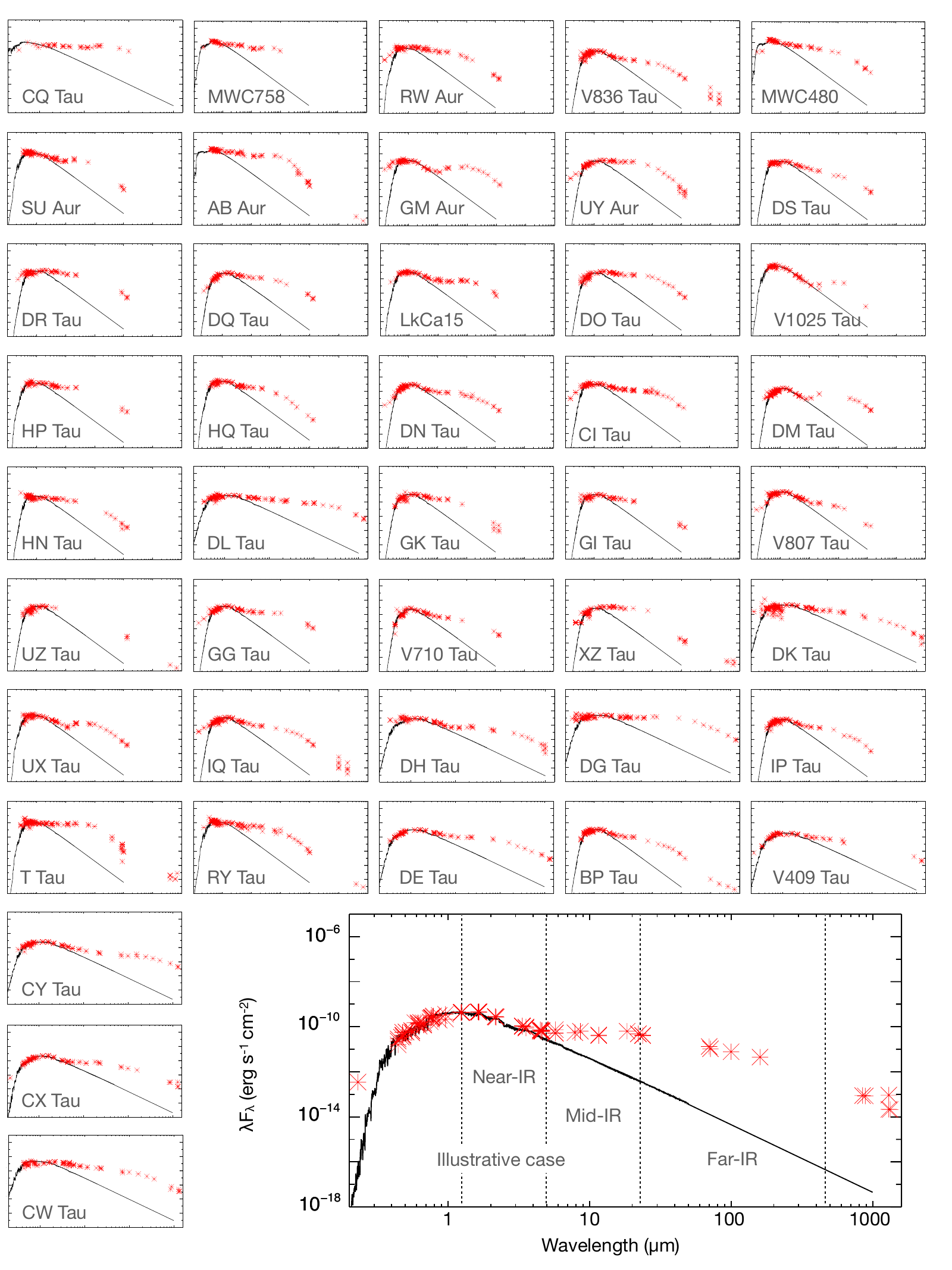} 
      \caption{Spectral energy distributions of entire sample. The de-reddened literature photometry (see Sect.\,\ref{sec:stellar_properties}) is indicated by the red symbols, while the \textit{\emph{Phoenix}} model for the stellar photosphere by is indicated
by the black line.}
          \label{fig:SEDs}
  \end{figure*}

The SEDs of the entire sample described in Sect.\,\ref{sec:stellar_properties} are shown in Fig.\,\ref{fig:SEDs}. In the following, we describe a few individual cases that require an in-depth analysis. The stellar and disk properties described in Sects.\,\ref{sec:stellar_properties} and \ref{sec:disk_properties} are listed in Tables \ref{table:stellar_sample} and \ref{table:disk_sample}.

Concerning HN Tau, there is no value of extinction, $A_{\rm V}$, scaled with the \citet{Cardelli1989} law that fits the entire optical photometry (0.3 to 0.9 $\mu$m). To avoid introducing the high level of uncertainty on the deriving properties, we did not include this source in the statistical analysis. In fact, even when scaling the model to the NIR photometry (less affected by the extinction), the resulting age calculated from isochrones turned out to be on the order of 10 Myr. In line with our finding, \citet{Alcala2021} defined this object as subluminous, explaining it with obscuration from the inclined disk \citep{Long2019}. We also note that this source is one of the eight with a very poor \textit{Gaia} solution (see Sect.\,\ref{sec:stellar_properties}), indicating clear problems in determining the photocenter. 

As noted in Sect.\,\ref{sec:overview}, V409 Tau and CW Tau are very variable in the visible. The origin of the variability is necessary to determine what level of observed photometry to adopt as the photospheric level. Here, we took advantage of the notion that their disk is very inclined (see Fig.\,\ref{fig:Imagery},  \citealt{Long2019}, and \citealt{Ueda2022}) to conclude that the variability is due to partial disk occultation and therefore adopt the brightest observed photometry as photospheric level. This means that the NIR photometry lies slightly below the stellar model, but this is probably due to the extinction still being  significant at these wavelengths.

\begin{table*}
\caption{Stellar properties of sample sorted by right ascension. Columns are target name, Taurus group following  \citet{Luhman2023}, distance, multiplicity, effective temperature, stellar luminosity, mass, and age.}             
\label{table:stellar_sample}      
\centering          
\begin{tabular}{l c c c c c c c}     
\hline\hline       
Target & Group & $d$ (pc) & System & $T_{\rm eff}$ (K) & $L_*$ (L$_{\odot}$) & $M_*$ (M$_{\odot}$) & $t$ (Myr) \\
\hline                    
   CQ Tau & HD35187$^*$ & 149.4 & (*) & 6900 & 6.30$\pm$0.27 & 1.4 & 12--14 \\  
   MWC758 & HD35187$^*$ & 155.9 & (*) & 8200 & 10.29$\pm$0.34 & 1.7 & 13--14 \\
   RW Aur & L1517 & 183.3$^*$ & (*)(*) & 5100+4400 & (1.4+0.7)$\pm$0.3 & 1.3+0.9 & 7.2--9.7 \\
   V836 Tau & L1544 & 167.0 & (*) & 3600 & 0.57$\pm$0.06 & 0.4 & 0.3--0.9 \\
   MWC480 & L1517 & 156.2 & (*) & 8800 & 19.90$\pm$0.65 & 2.1 & 6.3--6.6 \\
   SU Aur & L1517 & 157.0 & (*) & 5800 & 15.80$\pm$2.30 & 2.5 & 2.0--2.5 \\
   AB Aur & L1517 & 155.9 & (*) & 9800 & 40.58$\pm$0.80 & 2.4 & 3.9--4.4 \\
   GM Aur & L1517 & 158.1 & (*) & 4350 & 1.40$\pm$0.18 & 0.9 & 1.0--1.6 \\
   UY Aur & L1517 & 152.3 & (*)(*) & 3800+3600 & (0.65+0.33)$\pm$0.04 & 0.5+0.4 & 0.9--2.1 \\
   DS Tau & L1517 & 158.4 & (*) & 3900 & 0.93$\pm$0.10 & 0.5 & 0.7--1.0 \\
   DR Tau & L1558 & 193.0 & (*) & 4200 & 2.53$\pm$0.21 & 0.4 & 0.3--0.5 \\
   DQ Tau & L1558 & 195.4 & (**) & 3600 & (0.50+0.50)$\pm$0.04 & 0.4 & 0.7--1.2 \\
   EM$^*$ LkCa15 & L1536 & 157.2 & (*) & 4600 & 1.11$\pm$0.11 & 1.2 & 3.2--4.5 \\
   DO Tau & L1527 & 138.5 & (*) & 3800 & 0.56$\pm$0.06 & 0.5 & 1.0--1.4 \\
   V$^*$ V1025 Tau & L1536 & 167.3 & (*) & 5800 & 8.54$\pm$1.91 & 2.0 & 3.6--5.0 \\
   V$^*$ HP Tau & L1536 & 171.2 & (*) & 5000 & 1.87$\pm$0.61 & 1.5 & 3.5--4.1 \\
   HQ Tau & L1536 & 161.4$^*$ & (*) & 4900 & 3.24$\pm$0.72 & 1.3 & 1.5--1.9 \\
   DN Tau & L1524$^+$ & 128.6 & (*) & 3850 & 0.59$\pm$0.06 & 0.5 & 1.0--1.3 \\
   CI Tau & L1536 & 160.3 & (*) & 4600 & 1.35$\pm$0.26 & 1.0 & 1.3--2.1 \\
   DM Tau & L1551 & 144.0 & (*) & 3700 & 0.21$\pm$0.03 & 0.4 & 2.9--6.0 \\
   HN Tau & L1551 & 134.4$^*$ & (*)* & 4300 & - & (1.6) & - \\
   DL Tau & L1521$^+$  & 159.9 & (*) & 4000 & 0.89$\pm$0.11 & 0.6 & 0.9--1.2 \\
   GK Tau & L1524  & 129.1 & (*) & 4400 & 0.84$\pm$0.06 & 0.5 & 0.6--1.0 \\
   GI Tau & L1524 & 129.4 & (*) & 4100 & 0.80$\pm$0.11 & 0.7 & 1.5--1.7 \\
   V807 Tau & L1524 & 184.1$^*$ & (*)** & 4400+2*3600 & (2.2+2*0.9)$\pm$0.3 & 0.5+2*0.4 & 0.4--0.6 \\
   UZ Tau & L1524 & 123.1$^*$ & (**)** & 3700+3400 & (0.5+0.3)$\pm$0.09 & 0.4+0.3 & 0.9--1.4 \\
   GG Tau & L1551 & 145.5 & ((**)(*)) & 3800+2*3500 & (0.7+2*0.3)$\pm$0.05 & 0.5+2*0.4 & 0.8--2.0 \\
   V710 Tau & L1551 & 146.0 & (*)* & 3600 & 0.62$\pm$0.05 & 0.4 & 0.4--0.8 \\
   XZ Tau & L1551 & (144) & (*)(*) & 2*3600 & 2*(0.29$\pm$0.04) & 2*0.4 & 1.7--2.8 \\
   DK Tau & L1524 & 132.0 & (*)(*) & 3900+3600 & (1.2+0.5)$\pm$0.05 & 0.6+0.4 & 0.4--0.6 \\
   UX Tau & L1551 & 142.2 & (*)** & 5200+3700 & (1.8+0.4)$\pm$0.2 & 1.4+0.5 & 6.1--10.0 \\
   IQ Tau & L1524 & 131.5 & (*) & 3800 & 0.73$\pm$0.13 & 0.5 & 0.7--1.0 \\
   DH Tau & L1524 & 133.4 & (*)(*) & 3600 & 0.58$\pm$0.12 & 0.4 & 0.4--0.9 \\
   DG Tau & L1524 & 125.3$^*$ & (*) & 4000 & 1.53$\pm$0.29 & 0.6 & 0.3--0.6 \\
   IP Tau & HD28354 & 129.4 & (*) & 3800 & 0.41$\pm$0.03 & 0.6 & 1.8--2.8 \\
   T Tau & T Tau & 145.1 & (*)(**) & 5200 & 10.40$\pm$0.90 & 2+2+0.5 & 0.7--1.0 \\
   RY Tau & L1495 & 138.2$^*$ & (*) & 5700 & 11.80$\pm$2.90 & 2.3 & 2.4--2.6 \\
   DE Tau & L1495 & 128.0 & (*) & 3700 & 0.62$\pm$0.07 & 0.4 & 0.7--1.0 \\
   BP Tau & L1495 & 127.4 & (*) & 3700 & 0.69$\pm$0.04 & 0.4 & 0.5--1.0 \\
   V409 Tau & L1524 & 129.7 & (*) & 3600 & 0.22$\pm$0.03 & 0.4 & 2.7--5.6 \\
   CY Tau & L1495  & 126.3 & (*) & 3700 & 0.38$\pm$0.04 & 0.5 & 1.8--2.2 \\
   CX Tau & L1495 & 126.7 & (*) & 3500 & 0.26$\pm$0.02 & 0.4 & 1.3--2.8 \\
   CW Tau & L1495 & 131.5 & (*) & 4800 & 1.59$\pm$0.34 & 1.3 & 3.1--5.1 \\
\hline 
\end{tabular}
\tablefoot{The HD35187 group is formally an association near Taurus. The extended name of the group L1524 is L1524/L1529/B215 \citep{Luhman2023}, comprising the three individual clouds. The same applies to L1521 (L1521/B213) and L1495 (L1495/B209).  In sources with a poor \textit{Gaia} solution (see Sect.\,\ref{sec:spatial_distribution}), the distance is accompanied by an asterisk. In all other cases, the tabulated distance has an error of less than 1.5 pc. XZ Tau has no parallax available from \textit{Gaia,} and we assume the average distance to its group. Parentheses and asterisks indicate the geometry of disk and star(s), respectively. The effective temperature is relative to the primary and of any companion within 3\arcsec \ that is bright enough to account for a significant percentage of optical/NIR photometry (10\% of the primary). The masses of T Tau are obtained astrometrically \citep{Koehler2016} as most of the observed optical luminosity is from N (Sa and Sb are heavily extincted). The mass of HN Tau is obtained by \citet{Gangi2022} considering obscuration effects.}
\end{table*}

\begin{table*}
\caption{Disk properties of sample sorted by right ascension. Columns are NIR, MIR, and FIR excesses relative to the stellar luminosity, approximate disk dust mass, category of the SPHERE image as from Sect.\,\ref{sec:overview} measured contrast in scattered light, and reference publication of the SPHERE images shown in this work. {The observing programs are listed in Table \ref{table:observations}.}}             
\label{table:disk_sample}      
\centering          
\begin{tabular}{l c c c c c c l}     
\hline\hline       
Target & NIR (\%) & MIR (\%) & FIR (\%) & $M_{\rm dust}$ (M$_{\oplus}$) & Type & $\alpha_{\rm pol}$ & Reference \\
\hline                    
   CQ Tau & 26.8 & 28.3 & 24.3 & 108 & Bright & 6.4$\pm$1.2 & \citet{Hammond2022} \\  
   MWC758 & 33.6 & 9.4 & 13.7 & 49 & Bright & 3.1$\pm$0.3 & \citet{Benisty2015} \\
   RW Aur & 55.1 & 38.9 & 18.4 & 27 & Ambient & - & - \\
   V836 Tau & 6.4 & 6.2 & 4.1 & 24 & Faint & 1.8$\pm$0.6 & -\\
   MWC480 & 18.7 & 11.2 & 5.2 & 188 & Faint & 0.5$\pm$0.1 & - \\
   SU Aur & 6.4 & 5.8 & 3.7 & 9 & Bright & 10.1$\pm$0.7 & \citet{Ginski2021} \\
   AB Aur & 35.6 & 15.8 & 17.9 & 32 & Bright & 9.7$\pm$0.5 & \citet{Boccaletti2020} \\
   GM Aur & 0 & 5.7 & 21.2 & 130 & Bright & 16.4$\pm$1.8 & - \\
   UY Aur & 49.6 & 77.3 & 53.0 & 3 & Ambient & 3.5$\pm$0.3 & - \\
   DS Tau & 14.8 & 8.2 & 2.3 & 12 & Faint & <0.2 & - \\
   DR Tau & 47.8 & 44.6 & - & 128 & Bright & 3.0$\pm$0.6 & \citet{Mesa2022} \\
   DQ Tau & 14.7 & 18.3 & 12.0 & 79 & Faint & 0.4$\pm$0.2 & - \\
   EM$^*$ LkCa15 & 12.1 & 6.1 & 7.0 & 44 & Bright & 3.7$\pm$0.4 & \citet{Thalmann2016} \\
   DO Tau  & 78.0 & 71.0 & 56.0 & 62 & Ambient & 1.8$\pm$0.3 & \citet{Huang2022} \\
   V$^*$ V1025 Tau & 0 & 0 & 0.7 & 1 & Faint & <0.2 & - \\
   V$^*$ HP Tau & 39.3 & 19.8 & - & 45 & Ambient & - & - \\
   HQ Tau & 14.7 & 10.9 & 4.2 & 3 & Faint & 0.4$\pm$0.1 & - \\
   DN Tau & 9.2 & 8.9 & 7.9 & 41 & Faint & 0.3$\pm$0.3 & - \\
   CI Tau & 20.7 & 13.9 & 9.9 & 137 & Faint & 1.4$\pm$0.4 & \citet{Garufi2022b} \\
   DM Tau & 0 & 6.1 & 14.3 & 56 & Faint & 2.6$\pm$0.6 & - \\
   HN Tau & - & - & - & 8 & Faint & 1.7$\pm$1.5 & - \\
   DL Tau & 10.5 & 23.0 & 13.5 & 130 & Faint & 0.7$\pm$0.4 & - \\
   GK Tau & 30.6 & 22.0 & 8.4 & 2 & Faint & 0.3$\pm$0.3 & - \\
   GI Tau & 13.3 & 17.5 & - & 6 & Faint & 0.4$\pm$0.3 & - \\
   V807 Tau & 4.8 & 4.2 & 1.6 & 17 & Faint & <0.3 & - \\
   UZ Tau & 26.0 & - & - & 14 & Faint & 2.5$\pm$0.4 & Zurlo et al.\, in prep. \\
   GG Tau & 17.6 & 14.8 & 12.9 & 295 & Bright & 8.7$\pm$1.2 & \citet{Keppler2020} \\
   V710 Tau & 15.9 & 10.1 & 3.6 & 34 & Faint & 2.7$\pm$0.4 & - \\
   XZ Tau & 102 & 120 & 55.0 & 2 & Ambient & - & Zurlo et al.\, in prep. \\
   DK Tau & 16.1 & 12.2 & 6.4 & 9 & Faint & <0.2 & - \\
   UX Tau & 8.8 & 3.7 & 8.4 & 33 & Ambient & 8.7$\pm$0.9 & \citet{Menard2020} \\
   IQ Tau & 20.5 & 10.8 & 6.1 & 32 & Faint & 1.7$\pm$0.7 & - \\
   DH Tau & 17.6 & 6.1 & 5.6 & 1 & Faint & 0.9$\pm$0.2 & \citet{vanHolstein2021} \\
   DG Tau & 23.2 & 53.0 & 57.1 & 162 & Ambient & 6.5$\pm$0.8 & - \\
   IP Tau & 14.0 & 8.8 & 5.3 & 4 & Faint & 1.3$\pm$0.2 & - \\
   T Tau & 27.9 & 55.4 & 56.0 & 67 & Ambient & - & - \\
   RY Tau & 22.9 & 21.9 & 12.0 & 52 & Ambient & 3.8$\pm$0.6 & \citet{Valegard2022} \\
   DE Tau & 15.7 & 11.0 & 8.3 & 15 & Faint & 0.6$\pm$0.3 & - \\
   BP Tau & 13.7 & 9.8 & 5.6 & 20 & Faint & 0.5$\pm$0.3 & - \\
   V409 Tau & 7.5 & 15.6 & - & 9 & Faint & 2.9$\pm$0.5 & -\\
   CY Tau & 11.3 & 6.0 & 3.1 & 43 & Faint & 0.6$\pm$0.1 & - \\
   CX Tau & 10.6 & 13.3 & 7.1 & 5 & Faint & 0.4$\pm$0.3 & - \\
   CW Tau & 25.0 & 19.0 & 6.9 & 27 & Faint & 1.5$\pm$0.3 & - \\
\hline 
\end{tabular}
\tablefoot{Unreported excesses are due to uncertainty in the SED or poor available photometry. Unreported contrasts are due to the presence of ambient material that biases the measurement. All SPHERE datasets are in the H band except CQ Tau (J band), MWC758 (K), LkCa15 (J), and CY Tau (K).}
\end{table*}

\section{Observing setup} \label{appendix:setup}
{The setup of all the observations in this work is given in Table \ref{table:observations}. The individual detector integration time (DIT) spans from the 0.837 sec of the non-coronagraphic observations of V807 Tau to the 96 sec of DM Tau. However, most of the targets have been observed with a DIT of 32 sec or 64 sec.} 

{Instead, a larger variety of total integration times $t_{\rm int}$ is present, spanning this from 11 minutes for several sources from the GTO to $>$100 minutes for V710 Tau and UZ Tau. The adoption of the contrast $\alpha_{\rm pol}$ to assess the disk brightness allows us to minimize this diversity. In fact, the uncertainty on this value (tabulated in Table \ref{table:disk_sample}) only mildly depends on the $t_{\rm int}$. Sources with $\Delta \alpha_{\rm pol}\lesssim3\cdot10^{-4}$ were observed for an average of 46 minutes, those with $\Delta \alpha_{\rm pol}=3\cdot10^{-4}$ for 32 minutes, and those with $\Delta \alpha_{\rm pol}\gtrsim3\cdot10^{-4}$ for 21 minutes. Since the typical detection limit of a disk is $\alpha_{\rm pol}=3\cdot10^{-4}$ \citep[see also][]{Garufi2022b}, we conclude that the diverse integration times of the observations in this work do not have a significant impact on demographic results of this work. An illustrative example of this is CY Tau ($t_{\rm int}=51$ min) and CX Tau ($t_{\rm int}=11$ min). The former has a much better $\Delta \alpha_{\rm pol}$ ($10^{-4}$ vs $3\cdot10^{-4}$), but the $\alpha_{\rm pol}$ is comparable ($6\cdot10^{-4}$ vs $4\cdot10^{-4}$), meaning that the disk of CY Tau is observed with a better signal-to-noise ratio than that of CX Tau, but it is assessed to be equally faint.}

\begin{table*}
\caption{{Details of observations.}}             
\label{table:observations}      
\centering          
\begin{tabular}{l c c c c c c}     
\hline\hline       
   Target         & Program (PI)   & Filter    & Observing night      & DIT (sec)& N$_{\rm frames}$ & $t_{\rm int}$ (min)\\
   \hline                    
   CQ Tau         & 098.C-0760(B) (Benisty)  & J & 2017-10-06 & 50 & 40 & 33 \\  
   MWC758         & 106.21HJ.001 (Benisty)  &K$_{\rm s}$& 2020-12-25 & 16 & 140 & 37\\
   RW Aur         & 0104.C-0122(A) (Facchini)&    H      & 2019-12-04 & 16 &  192 & 51\\
   V836 Tau       & 0102.C-0453(A) (GTO, Beuzit)&    H      & 2019-09-01 & 32 &  20 & 11\\
   MWC480         & 0101.C-0867(A) (Keppler) &    H      & 2018-11-17 & 32  &  108 &  57\\
   SU Aur         & 1104.C-0415(E) (DESTINYS, Ginski)&    H      & 2019-12-14 & 32 & 104 & 55\\
   AB Aur         & 0104.C-0157(B) (Boccaletti) &    H      & 2019-12-18 & 32 &  176 & 94\\
   GM Aur         & 0100.C-0452(A) (Benisty) &    H      & 2018-01-02 & 64 &  32 & 34\\
   UY Aur         & 1104.C-0415(G) (DESTINYS, Ginski)&    H      & 2021-01-21 & 16 &  94 & 25\\
   DS Tau         & 0102.C-0453(A) (GTO, Beuzit) &    H      & 2018-12-25 & 32 &  20 & 11\\
   DR Tau         & 0102.C-0453(A) (GTO, Beuzit) &    H      & 2018-11-25 & 32 &  20 & 11\\
   DQ Tau         & 1104.C-0415(E) (DESTINYS, Ginski)&    H      & 2019-11-24 & 64 &  56 & 60\\
   EM$^*$ LkCa15  & 096.C-0248(A) (GTO, Beuzit)  &    J      & 2020-12-07 & 32 & 120 & 64 \\
   DO Tau         & 1104.C-0415(E) (DESTINYS, Ginski)&    H      & 2019-12-19 & 64 &  56 & 60\\
   V$^*$ V1025 Tau& 1104.C-0415(A) (DESTINYS, Ginski)&    H      & 2019-12-15 & 64 &  56 & 60\\
   V$^*$ HP Tau   & 1104.C-0415(E) (DESTINYS, Ginski)&    H      & 2019-12-19 & 64 &  56 & 60\\
   HQ Tau         & 0102.C-0453(A) (GTO, Beuzit) &    H      & 2018-11-28 & 32 &  20 & 11\\
   DN Tau         & 0102.C-0453(A)  (GTO, Beuzit) &    H      & 2019-08-15 & 32 &  20 & 11\\
   CI Tau         & 0100.C-0452(A)  (Benisty) &    H      & 2018-01-01 & 64 &  32 & 34\\
   DM Tau         & 0101.C-0867(A) (Keppler) &    H      & 2018-10-01 & 96 &  36 & 58\\
   HN Tau         & 1104.C-0415(E) (DESTINYS, Ginski)&    H      & 2019-12-13 & 64 &  12 & 13\\
   DL Tau         & 0102.C-0453(A) (GTO, Beuzit) &    H      & 2018-12-28 & 32 &  20 & 11\\
   GK Tau         & 0102.C-0453(A) (GTO, Beuzit)&    H      & 2019-10-24 & 32 &  20 & 11\\
   GI Tau         & 0102.C-0453(A) (GTO, Beuzit)&    H      & 2018-11-26 & 32 &  20 & 11\\
   V807 Tau       & 1104.C-0415(E) (DESTINYS, Ginski)&    H      & 2019-12-30 &0.837&4320 &  60\\
   UZ Tau         & 0100.C-0408(B) (Cieza) &    H      & 2017-10-27 & 64 &  108 & 115 \\
   GG Tau         &  198.C-0209(N) (Beuzit) &    H      & 2016-11-18 &  4 &  660 &  44 \\ 
   V710 Tau       & 1104.C-0415(A) (DESTINYS, Ginski)&    H      & 2019-12-20 &  4 &  1536 &  102\\
   XZ Tau         & 0100.C-0408(A) (Cieza) &    H      & 2017-10-12 & 64 &  90 & 96\\
   DK Tau         & 0102.C-0453(A) (GTO, Beuzit) &    H      & 2019-09-01 & 32 &  20 & 23\\
   UX Tau         & 0100.C-0452(B) (Benisty) &    H     & 2017-10-05 &  64 & 48 & 51 \\
   IQ Tau         & 0102.C-0453(A) (GTO, Beuzit) &    H      & 2018-12-18 & 32 &  44 & 23\\
   DH Tau         & 0102.C-0916(A) (GTO, Beuzit) &    H      & 2019-10-23 & 64 &  36 & 38\\
   DG Tau         & 1104.C-0415(D) (DESTINYS, Ginski)&    H      & 2021-10-28 & 64 &  56 & 60\\
   IP Tau         & 1104.C-0415(A) (DESTINYS, Ginski)&    H      & 2019-12-14 & 64 &  56 & 60\\
   T Tau          &  198.C-0209(N) (GTO, Beuzit) &    H      & 2016-11-18 & 16 &  64 &  17\\
   RY Tau         & 1104.C-0415(E) (DESTINYS, Ginski)&    H      & 2018-12-16 & 32 &  68 & 36\\
   DE Tau         & 0102.C-0453(A) (GTO, Beuzit) &    H      & 2019-10-11 & 32 &  20 &  11\\
   BP Tau         & 0102.C-0453(A) (GTO, Beuzit) &    H      & 2018-12-22 & 32 &  20 & 11\\
   V409 Tau       & 1104.C-0415(A) (DESTINYS, Ginski)&    H      & 2019-11-16 & 64 &  56 & 60\\
   CY Tau         &  108.22EE.001 (Benisty) &K$_{\rm s}$& 2021-12-26 & 32 &  96 & 51\\    
   CX Tau         & 0102.C-0453(A) (GTO, Beuzit) &    H      & 2019-10-23 & 32 &  20 & 11\\
   CW Tau         & 0102.C-0453(A) (GTO, Beuzit) &    H      & 2018-12-22 & 32 &  32 & 17\\    
\hline 
\end{tabular}
\tablefoot{{DIT is the detector integration time, N$_{\rm frames}$ is the number of individual frames, $t_{\rm int}$ is the total on-target integration time ($t_{\rm obs}={\rm DIT}\times{\rm NDIT}$).}}
\end{table*}

\section{Notes on SPHERE images with no prior publications} \label{appendix:individual}
Here, we briefly describe the SPHERE images with no prior publications available. Sources that are not mentioned are published as referenced in Table \ref{table:disk_sample}. {The basic constraints on the disk geometry extractable from the SPHERE images (outermost separation with detectable signal, crude inclination, and position angle) are listed in Table \ref{table:disk_geometry}.} In the following, we refer to a $QU$ pattern as the expected alignment of positive and negative signal in the $Q$ and $U$ image in case of centro-symmetric scattering, and that can be used to infer a disk detection where the signal detected in the $Q_\phi$ image is of dubious origin \citep[see e.g., Appendix C in][]{Garufi2022b}. Furthermore, the disk extent constrained by ALMA is from the continuum emission unless otherwise specified. 

\begin{table}
\caption{{Disk geometry interpreted from the SPHERE image.}}             
\label{table:disk_geometry}      
\centering          
\begin{tabular}{l c c c}     
\hline\hline  
Target & $r_{\rm out}$ (\arcsec) & $i$ & P.A. ($\degree$) \\ 
\hline
CQ Tau & 0.6 & Intermediate & 45 \\
MWC758  & 0.6 & Mild & - \\
RW Aur   & 0.5 & Intermediate & 45 \\
V836 Tau   & 0.2 & Intermediate & 120 \\
MWC480   & 1.0 & Intermediate & 50 \\
SU Aur   & 0.5 & Mild & - \\
AB Aur   & 2.4 & Intermediate & 45 \\
GM Aur   & 2.0 & Intermediate & 60 \\
UY Aur   & - & - & - \\
DS Tau   & - & - & - \\
DR Tau   & 0.7 & - & - \\
DQ Tau   & 0.2 & Intermediate & 20 \\
LkCa15   & 0.7 & Intermediate & 60 \\
DO Tau   & 0.6 & - & - \\
V1025 Tau   & - & - & - \\
HP Tau   & - & - & - \\
HQ Tau   & 0.2 & - & - \\
DN Tau   & 0.6 & - & - \\
CI Tau   & 0.5 & Intermediate & 10 \\
DM Tau   & 0.4 & Intermediate & 90 \\
HN Tau   & - & - & - \\
DL Tau   & 0.8 & Intermediate & 50 \\
GK Tau   & - & - & - \\
GI Tau   & 0.2 & - & - \\
V807 Tau   & - & - & - \\
UZ Tau   & 0.6 & High & 90 \\
GG Tau   & 1.8 & Intermediate & 110 \\
V710 Tau   & 0.4 & Intermediate & 70 \\
XZ Tau   & - & - & - \\
DK Tau   & - & - & - \\
UX Tau   & 0.6 & Intermediate & 80 \\
IQ Tau   & 0.5 & High & 45 \\
DH Tau   & 0.2 & - & - \\
DG Tau   & 0.7 & Intermediate & 140 \\
IP Tau   & 0.3 & Intermediate & 70 \\
T Tau   & - & - & - \\
RY Tau   & - & - & - \\
DE Tau   & 0.2 & - & - \\
BP Tau   & 0.3 & Intermediate & 150 \\
V409 Tau   & 0.5 & High & 40 \\
CY Tau   & 0.2 & - & - \\
CX Tau   & 0.4 & - & - \\
CW Tau   & 0.4 & High & 70 \\
\hline 
\end{tabular}
\tablefoot{Columns are target, outermost separation with detectable signal, indicative inclination, and position angle.}
\end{table}

\textit{RW Aur}. The $Q_\phi$ image reveals a peculiar morphology around the primary star. The $QU$ pattern indicates the detection of a disk with position angle of approximately 45\degree. However, bright signal extended in the direction orthogonal to the disk is also detected, and this likely probes the two outflow cavities in a morphology similar to RY Tau \citep{Garufi2019}. Interestingly, a narrow dark lane resembling a gap sculpted by a collimated jet is visible to the SE. Since the disk signal is blended with the outflow material (see Sect.\,\ref{sec:ambient}), the calculation of $\alpha_{\rm pol}$ is not provided in this work.

\textit{V836 Tau}. A small disk of moderate brightness (with a $\alpha_{\rm pol}$ of 1.8$\pm$0.6) is clearly visible in the $Q_\phi$ image. The apparent disk extent (0.2\arcsec) is comparable to ALMA \citep{Long2019}.

\textit{MWC480}. As discussed in Sects.\,\ref{sec:faint} and \ref{sec:ALMA_shadowed}, this disk is faint given the stellar luminosity, but some substructures are immediately visible from the image. Beside the central component and the first ring at 0.6\arcsec, a second ring consistent with the disk geometry is marginally visible to the SE 1\arcsec\ from the star. 

\begin{figure*}
  \sidecaption
  \includegraphics[width=12cm]{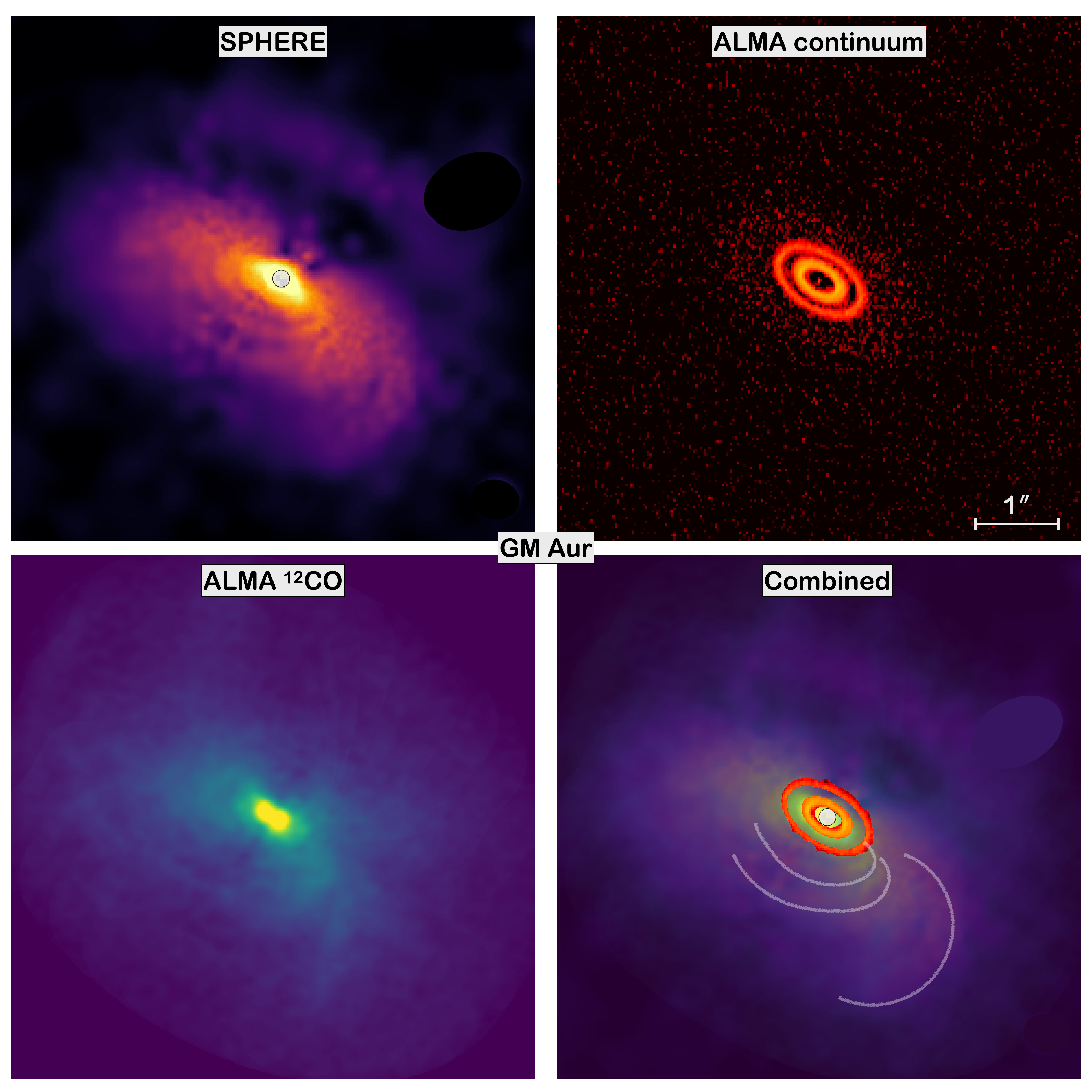} 
  \caption{Multicomponent view of GM Aur. \textit{Top left}: SPHERE image from this work obtained with an adaptive de-noising procedure based on the adaptive kernel smoothing technique and implemented in the \textsc{splash} software \citep{Price2007}. \textit{Top right}: ALMA continuum image in Band 4 from \citet{Huang2020}. \textit{Bottom left}: ALMA $^{12}$CO 2--1 line moment-0 map from \citet{Huang2021}. \textit{Bottom right}: Combined view. The most obvious arms are highlighted in this panel. The common spatial scale is indicated to the bottom right.}
          \label{fig:GM_Aur}
\end{figure*}

\textit{GM Aur}. This disk is among the few most spectacular detections of the whole sample. The relatively inclined disk \citep[53\degree\ from the ALMA image by][]{Huang2020} is visible from both the front and rear disk surfaces that are well separated by a geometrically thick dark lane with an apparent size of nearly 0.5\arcsec. Along the disk major axis, the disk is detected out to 2\arcsec, that is twice as large as the ALMA continuum extent \citep{Huang2020}. A number of spiral-like structures are visible from the disk front surface as highlighted in Fig.\,\ref{fig:GM_Aur}. The $^{12}$CO 2--1 line moment-0 map from \citet{Huang2021} shows spiral arms with evident spatial analogies with the SPHERE image.

\textit{UY Aur}. The polarimetric image shows a number of arc-like structures that are reasonably due to the interaction between the binaries. Given the very low disk extent constrained by ALMA \citep[6 au,][]{Long2019}, it is probable that none of the detected signal originates from the disk of either component. The $Q_\phi$ image shows some strong negative signal due to the complex summation of the individual stellar contributions.

\textit{DS Tau}. No $QU$ pattern is visible. Nonetheless, some $Q_\phi$ signal that is spatially consistent with the ALMA image is visible from Fig.\,\ref{fig:shadowed}. A SPHERE image at longer wavelengths (K band) reveals a relatively clear detection (Ren et al.\,in prep.). Yet, in this work this is treated as a non-detection.

\textit{DQ Tau}. A relatively clear disk is detected out to 0.2\arcsec, which is the extent of the ALMA continuum emission by \citet{Long2019}. Nonetheless, this disk is rather massive in terms of dust (79 M$_\oplus$), making it a good example of a self-shadowed disk (see Fig.\,\ref{fig:trends}).

\textit{V* V1025 Tau}. No $QU$ pattern is visible. It is treated as a non-detection. As discussed throughout the text and shown in Fig.\,\ref{fig:trends}, it is a Class III object. It is very close in space to HP Tau (see next source).

\textit{V* HP Tau}. A very complex pattern of arc-like structures is visible from the $Q_\phi$ image. The brightest arcs are distributed within the inner 2\arcsec,\ but some fainter structures are visible out to the detector edge at 6\arcsec. The disk has an appreciable dust mass (45 M$_\oplus$ from our calculation). No companion that may be responsible for the peculiar features observed is visible from the SPHERE image, but the source is close to CoKu HP Tau G3 (17\arcsec) and V1025 Tau (21\arcsec; see previous source), as well as to the bright reflection nebula, GN 04.32.8.

\textit{HQ Tau}. A very faint ($\alpha_{\rm pol}$=0.4$\pm$0.1) and compact (0.2\arcsec) $QU$ pattern is visible, suggesting a vertically oriented disk that is consistent with the ALMA image by \citet{Long2019}. With only 3 M$_\oplus$ estimated, this is one of the lowest mass disks ever resolved in the NIR (see Fig.\,\ref{fig:trends}).

\textit{DN Tau}. Despite the very low contrast measured (0.3$\pm$0.3), a faint $QU$ pattern is visible out to 0.6\arcsec. As shown in Fig.\,\ref{fig:shadowed}, it is a prototypical example of a self-shadowed disk.

\textit{DM Tau}. The detection of the faint disk of DM Tau is described in depth in Sect.\,\ref{sec:ALMA_cavity}. Readers may refer to that.

\textit{HN Tau}. The AO correction is imperfect because of the low stellar luminosity (G=13.4; see Appendix \ref{appendix:sample}). It is treated as a marginal detection because the $Q_\phi$ signal aligns with the disk position angle by \citet{Long2019}. However, the calculated $\alpha_{\rm pol}$ and relative error ($1.7\pm$1.5) reflects this uncertainty.

\textit{DL Tau}. Despite the low contrast measured (0.7$\pm$0.4), the disk is resolved out to 0.8\arcsec. As shown in Fig.\,\ref{fig:shadowed}, it is a prototypical example of self-shadowed disk where some disk substructures are visible after careful inspection.

\textit{GK Tau}. No $QU$ pattern is visible. It is treated as a non-detection. The low dust mass calculated (2 M$_\oplus$) suggests a small disk \citep[<9 au in size from][]{Long2019}.

\textit{GI Tau}. A very faint ($\alpha_{\rm pol}$=0.4$\pm$0.3) and compact (0.2\arcsec) $QU$ pattern is visible. Similarly to HQ Tau, with only 6 M$_\oplus$ estimated, this is one of the lowest mass disks ever resolved in the NIR.

\textit{V807 Tau}. The SPHERE image is non-coronagraphic. The binary companion Bab is visible from the image at 0.17\arcsec\ and is in turn marginally resolved. Owing to the presence of the companion, no $QU$ pattern can be identified. It is treated as a non-detection. The ALMA image by \citet{Akeson2019} reveals a marginally resolved emission where the individual components of A and Bab are not resolved.

\textit{UZ Tau}. A clear, moderately inclined disk is visible from the $Q_\phi$ image. The measured contrast (2.5$\pm$0.4) is close to the formal limit between bright and faint disks adopted in this work. As shown in Fig.\,\ref{fig:shadowed}, it is a prototypical example of self-shadowed disk. This source will be studied in a forthcoming publication (Zurlo et al.)

\textit{V710 Tau}. The clear disk detected in the $Q_\phi$ image has a dip with negative values to the north. Along the disk major axis, the disk appears much more extended to the east (0.4\arcsec) than to the west (0.25\arcsec). We are inclined to consider these effects as real, and not due to an imperfect centering or stellar polarization subtraction.

\textit{XZ Tau}. Multiple extended structures that are at least partly due to the interaction between the binaries are visible in the polarimetric image. This source will be studied in a forthcoming publication (Zurlo et al.).

\textit{DK Tau}. No $QU$ pattern is visible. It is treated as a non-detection. The low dust mass calculated (9 M$_\oplus$) suggests a small disk \citep[<15 au in size from][]{Long2019}.

\textit{IQ Tau}. A faint yet extended (0.5\arcsec) disk is visible from the $Q_\phi$ image, with an extent comparable to ALMA (see Fig.\,\ref{fig:shadowed}). 

\textit{DG Tau}. The $Q_\phi$ image of DG Tau reveals a  complicated field with a prominent contribution from the natal envelope and from extended arms that will be studied in a forthcoming work.

\textit{IP Tau}. The detection of the faint disk of IP Tau is described in depth in Sect.\,\ref{sec:ALMA_cavity}. Readers may refer to that.

\textit{T Tau}. In line with previous work \citep[e.g.,][]{Kasper2020}, the $Q_\phi$ image of T Tau reveals an extremely complicated field, with multiple arms, arcs, and dips that will be studied in a forthcoming publication (van Holstein et al.).

\textit{DE Tau}. The $Q_\phi$ image only reveals faint signal along the SE-NW axis from the inner 0.15\arcsec.

\textit{BP Tau}. Similarly to V710 Tau, the disk that is clearly visible from the $Q_\phi$ image exhibits a dip to the north and an obvious azimuthal asymmetry, with the SE portion of the disk being brighter than the NW region. 

\textit{V409 Tau}. A clear, inclined disk is visible from the $Q_\phi$ image in line with the ALMA image \citep[$i=69\degree$,][]{Long2019}. The measured contrast (2.9$\pm$0.5) is at the formal limit between bright and faint disks adopted in this work. However, it is known that inclined disks appear artificially brighter in scattered light because we inevitably probe the forward peak of the phase function \citep{Garufi2022b}.

\textit{CY Tau}. The $Q_\phi$ image reveals a faint emission with an approximate position angle of 160\degree\ pointing to a relatively inclined disk that is nearly vertically oriented. This possible morphology is coarsely consistent with the archival ALMA data from program 2013.1.00498 (PI: Perez, L.).

\textit{CX Tau}. The presence of a faint $QU$ pattern can be inferred out to 0.35\arcsec. The presence of a gaseous disk more extended than 75 au \citep{Facchini2019} surrounding a compact continuum emission (14 au) indicates that this is also a self-shadowed disk.   

\textit{CW Tau}. A clearly inclined disk is inferred from the $Q_\phi$ image that is geometrically consistent with the ALMA image by \citet{Ueda2022}.

\end{appendix}

\end{document}